\documentclass[final]{siamltex}
\usepackage{waveletnotation}
\usepackage{epsfig}

\newcommand{\e}{\mathbf e}
\newcommand{\mrow}{r}
\newcommand{\mcol}{s}
\newcommand{\cs}{\mbox{\rm{coeffsupp}}}
\newcommand{\coeff}{\mbox{\rm{coeff}}}
\newcommand{\sgn}{\mbox{\rm{sgn}}}
\renewcommand{\PP}{\mathcal{P}}

\newtheorem{example}{Example}
\newtheorem{algorithm}{Algorithm}

\begin{document}

\title{Matrix Extension with Symmetry and its Application to Filter Banks\thanks{Research supported in part by NSERC Canada under Grant RGP 228051.}}

\author{Bin Han\thanks{Department of Mathematical and Statistical Sciences, University of Alberta, Edmonton, \quad Alberta, Canada T6G 2G1. {\tt bhan@math.ualberta.ca}, {\tt xzhuang@math.ualberta.ca}\quad {\tt http://www.ualberta.ca/$\sim$bhan, http://www.ualberta.ca/$\sim$xzhuang}
} \and Xiaosheng Zhuang\footnotemark[2]}

\maketitle

\makeatletter \@addtoreset{equation}{section} \makeatother

\begin{abstract}
Let $\pP$ be an $r \times s$ matrix of Laurent polynomials with
symmetry such that $\pP(z) \pP^*(z)=I_\mrow$ for all $z\in \C \bs
\{0\}$ and the symmetry of $\pP$ is compatible. The matrix extension
problem with symmetry is to find an $s \times s$ square matrix
$\pP_e$ of Laurent polynomials with symmetry such that $[I_r,
\mathbf{0}] \pP_e =\pP$ (that is, the submatrix of the first $r$
rows of $\pP_e$ is the given matrix $\pP$), $\pP_e$ is paraunitary
satisfying $\pP_e(z)\pP_e^*(z)=I_\mcol$ for all $z\in \C \bs \{0\}$,
and the symmetry of $\pP_e$ is compatible. Moreover, it is highly
desirable in many applications that the support of the coefficient
sequence of $\pP_e$ can be controlled  by that of $\pP$. In this
paper, we
 completely solve the matrix extension problem with symmetry
and provide a step-by-step algorithm to construct such a desired
matrix $\pP_e$ from a given matrix $\pP$. Furthermore, using a
cascade structure, we obtain a complete representation of any
$r\times s$ paraunitary matrix $\pP$ having compatible symmetry,
which in turn leads to an algorithm for deriving a desired matrix
$\pP_e$ from a given matrix $\pP$. Matrix extension plays an
important role in many areas such as electronic engineering, system
sciences, applied mathematics, and pure mathematics. As an
application of our general results on matrix extension with
symmetry, we obtain a satisfactory algorithm for constructing
symmetric paraunitary filter banks and symmetric orthonormal
multiwavelets by deriving high-pass filters with symmetry from any
given low-pass filters with symmetry. Several examples are provided
to illustrate the proposed algorithms and results in this paper.
\end{abstract}



\bigskip

\pagenumbering{arabic}

\section{Introduction and Main Results}

The matrix extension problem plays a fundamental role in many areas
such as electronic engineering, system sciences, mathematics, and
etc. To mention only a few references here on this topic, see
\cite{CL,Cui,Daub:book,
GHM,Han:d4,Han:jfaa:2009,HJ:d4,JS,Jiang2,LLS,P,Shen,V,YP}. For
example, matrix extension is an indispensable tool in the design of
filter banks  in electronic engineering (\cite{Jiang1,Jiang2,V,YP})
and in the construction of multiwavelets in wavelet analysis
(\cite{CL,Cui,Daub:book,DHardin,GHM,Han:d4,Han:jfaa:2009,HJ:d4,HanMo,JS,LLS,Turca,P}).
In order to state the matrix extension problem and our main results
on this topic, let us introduce some notation and definitions first.

Let $\pp(z)=\sum_{k\in \Z} p_k z^k, z\in \C \bs \{0\}$ be a Laurent polynomial with complex coefficients $p_k\in \C$ for all $k\in \Z$.
We say that $\pp$ has {\it symmetry} if its coefficient sequence $\{p_k\}_{k\in \Z}$ has symmetry; more precisely, there exist $\gep \in \{-1,1\}$ and $c\in \Z$ such that
\begin{equation}\label{sym:seq}
p_{c-k}=\gep p_k, \qquad \forall\; k\in \Z.
\end{equation}
If $\gep=1$, then $\pp$ is symmetric about the point $c/2$; if $\gep=-1$, then $\pp$ is antisymmetric about the point $c/2$.
Symmetry of a Laurent polynomial can be conveniently expressed using a symmetry operator $\sym$ defined by
\begin{equation}\label{sym}
\sym \pp(z):=\frac{\pp(z)}{\pp(1/z)}, \qquad z\in \C \bs \{0\}.
\end{equation}
When $\pp$ is not identically zero,  it is evident that
\eqref{sym:seq} holds if and only if $\sym \pp(z)=\gep z^c$. For the
zero polynomial, it is very natural that $\sym 0$ can be assigned
 any symmetry pattern; that is, for every occurrence
of $\sym 0$ appearing in an identity in this paper, $\sym 0$  is
understood to take an appropriate choice of $\gep z^c$ for some
$\gep\in \{-1,1\}$ and $c\in \Z$ so that the identity holds. If
$\pP$ is an $r\times s$ matrix of Laurent polynomials with symmetry,
then we can apply the operator $\sym$ to each entry of $\pP$, that
is, $\sym \pP$ is an $r\times s$ matrix such that $[\sym
\pP]_{j,k}:=\sym ([\pP]_{j,k})$, where $[\pP]_{j,k}$ denotes the
$(j,k)$-entry of the matrix $\pP$ throughout the paper.

For two matrices $\pP$ and $\pQ$ of Laurent polynomials with
symmetry, even though all the entries in $\pP$ and $\pQ$ have
symmetry, their sum $\pP+\pQ$, difference $\pP-\pQ$, or product
$\pP\pQ$, if well defined, generally may not have symmetry any more.
This is one of the difficulties for matrix extension with symmetry.
In order for $\pP\pm\pQ$ or $\pP \pQ$ to possess some symmetry, the
symmetry patterns of $\pP$ and $\pQ$ should be compatible. For
example, if $\sym \pP=\sym \pQ$, that is, both $\pP$ and $\pQ$ have
the same symmetry pattern, then indeed $\pP\pm\pQ$ has symmetry and
$\sym(\pP\pm\pQ)=\sym \pP=\sym \pQ$.  In the following, we discuss
the compatibility of symmetry patterns of matrices of Laurent
polynomials. For an $r\times s$ matrix $\pP(z)=\sum_{k\in \Z} P_k
z^k$, throughout the paper we denote
\begin{equation}\label{Pstar}
\pP^*(z):=\sum_{k\in \Z} P_k^* z^{-k} \quad \mbox{with}
%\quad z^*:=z^{-1} \quad \mbox{and}
\quad P_k^*:=\ol{P_k}^T,\qquad k\in \Z,
\end{equation}
where $\ol{P_k}^T$ denotes the transpose of the complex conjugate of
the constant matrix $P_k$ in $\C$. We say that \emph{the symmetry of
$\pP$ is compatible} or \emph{$\pP$ has compatible symmetry}, if
\begin{equation}\label{sym:comp}
\sym \pP(z)=(\sym \pth_1)^*(z) \sym \pth_2(z),
\end{equation}
for some $1 \times r$ and $1\times s$ row vectors $\pth_1$ and
$\pth_2$ of Laurent polynomials with symmetry. For an $r\times s$
matrix $\pP$ and an $s\times t$ matrix $\pQ$ of Laurent polynomials,
we say that $(\pP, \pQ)$  \emph{has mutually compatible symmetry} if
%the symmetry patterns of $\pP$ and $\pQ$ are
%\emph{mutually compatible} if
%
\begin{equation}\label{sym:mutual}
\sym \pP(z)=(\sym \pth_1)^*(z) \sym \pth(z) \quad \mbox{and} \quad
\sym \pQ(z)=(\sym \pth)^*(z) \sym \pth_2(z)
\end{equation}
for some $1\times r$, $1\times s$,  $1\times t$ row vectors $\pth_1,
\pth, \pth_2$ of Laurent polynomials with symmetry. If $(\pP,\pQ)$
has mutually compatible symmetry as in \eqref{sym:mutual}, then it
is easy to verify that their product $\pP \pQ$ has compatible
symmetry and in fact $\sym (\pP\pQ)= (\sym \pth_1)^*\sym \pth_2 $.

For a matrix of Laurent polynomials, another important property is
the support of its coefficient sequence. For $\pP=\sum_{k\in \Z} P_k
z^k$ such that $P_k={\bf0}$ for all $k\in \Z \bs [m,n]$ with
$P_{m}\ne{\bf0}$ and $P_{n}\ne {\bf0}$, we define its coefficient
support to be $\cs(\pP):=[m,n]$ and the length of its coefficient
support to be $|\cs(\pP)|:=n-m$. In particular, we define
$\cs({\bf0}):=\emptyset$, the empty set, and
$|\cs({\bf0})|:=-\infty$. Also, we  use $\coeff(\pP,k):=P_k$ to
denote the coefficient matrix (vector) $P_k$ of $z^k$ in $\pP$. In
this paper, ${\bf0}$ always denotes a general zero matrix whose size
can be determined in the context.

The Laurent polynomials that we shall consider in this paper have
their coefficients in a subfield $\F$ of the complex field $\C$. Let
$\F$ denote a subfield of $\C$ such that $\F$ is closed under the
operations of complex conjugate of $\F$ and square roots of positive
numbers in $\F$. In other words, the subfield $\F$ of $\C$ satisfies
the following properties:
\begin{equation}\label{F}
\bar{x}\in \F \quad \hbox{and}\quad \sqrt{y}\in \F, \qquad \forall\;
x,y \in \F\quad \mbox{with}\quad y>0.
\end{equation}
Two particular examples of such subfields $\F$ are $\F=\R$ (the
field of real numbers) and $\F=\C$ (the field of complex numbers).

Now, we introduce the general matrix extension problem with
symmetry. Throughout the paper,   $r$ and $s$ denote two positive
integers such that $1\le r\le s$. Let $\pP$ be an $r\times s$ matrix
of Laurent polynomials with coefficients in $\F$ such that $\pP(z)
\pP^*(z)=I_{\mrow}$ for all $z\in \C \bs \{0\}$ and the symmetry of
$\pP$ is compatible, where $I_r$ denotes the $r\times r$ identity
matrix. The matrix extension problem with symmetry is to find an
$s\times s$ square matrix $\pP_e$ of Laurent polynomials with
coefficients in $\F$ and with symmetry such that $[I_r, \mathbf{0}]
\pP_e=\pP$ (that is, the submatrix of the first $r$ rows of $\pP_e$
is the given matrix $\pP$), the symmetry of $\pP_e$ is compatible,
and $\pP_e(z)\pP_e^*(z)=I_{\mcol}$ for all $z\in \C \bs \{0\}$ (that
is, $\pP_e$ is paraunitary).  Moreover, in many applications, it is
often highly desirable that the coefficient support of $\pP_e$ can
be controlled by that of $\pP$ in some way.

In this paper, we study this general matrix extension problem with symmetry and we completely solve this problem as follows:

\begin{theorem}\label{thm:main:1}
Let $\F$ be a subfield of $\C$ such that \eqref{F} holds. Let $\pP$
be an $r\times s$ matrix of Laurent polynomials with coefficients in
$\F$ such that the symmetry of $\pP$ is compatible and
$\pP(z)\pP^*(z)=I_r$ for all $z\in \C \bs \{0\}$. Then there exists
an $s \times s$ square matrix $\pP_e$, which can be constructed by
Algorithm~\ref{alg:1} in section~2 from the given matrix $\pP$, of
Laurent polynomials with coefficients in $\F$ such that
\begin{enumerate}
\item[{\rm(i)}] $[I_r, \mathbf{0}] \pP_e=\pP$, that is, the submatrix of the first $r$ rows of $\pP_e$ is $\pP$;
\item[{\rm(ii)}] $\pP_e$ is paraunitary: $\pP_e(z) \pP_e^*(z)=I_s$ for all $z\in \C \bs \{0\}$;
\item[{\rm(iii)}] The symmetry of $\pP_e$ is compatible;
\item[{\rm(iv)}] The coefficient support of $\pP_e$ is controlled by that of $\pP$ in the following
sense:
\begin{equation}\label{supp:control}
|\cs([\pP_e]_{j,k})|\le \max_{1\le n\le r} |\cs([\pP]_{n,k})|,
\qquad 1\le j, k\le s.
\end{equation}
%i.e., the length of the coefficient support of any entry in the
%$j$-th column of $\pP_e$ is
%controlled by that of the entry in the $j$-th column of $\pP$ with  maximal length of coefficient support.%
\end{enumerate}
\end{theorem}

Theorem~\ref{thm:main:1} on matrix extension with symmetry is built
on a stronger result which represents any given paraunitary matrix
having compatible symmetry by a simple cascade structure. The
following result leads to a proof of Theorem~\ref{thm:main:1} and
completely characterizes any paraunitary matrix $\pP$ in
Theorem~\ref{thm:main:1}.

\begin{theorem}\label{thm:main:2}
Let $\pP$ be an $r\times s$ matrix of Laurent polynomials with
coefficients in a subfield $\F$ of $\C$ such that \eqref{F} holds. Then
$\pP(z) \pP^*(z)=I_r$ for all $z\in \C \bs \{0\}$
%(that is, $\pP$ is paraunitary)
and the symmetry of $\pP$ is compatible as in \eqref{sym:comp}, if
and only if, there exist $s\times s$ matrices $\pP_0, \ldots,
\pP_{J+1}$ of Laurent polynomials with coefficients in $\F$ such
that
\begin{enumerate}
\item[{\rm(1)}] $\pP$ can be represented as a product of  $\pP_0,\ldots,\pP_{J+1}$:
\begin{equation}\label{cascade}
\pP(z)=[I_r,\mathbf{0}] \pP_{J+1}(z) \pP_J(z)\cdots \pP_1(z)\pP_0(z);
\end{equation}

\item[{\rm(2)}] $\pP_j, 1\le j\le J$ are elementary: $\pP_j(z)\pP_j^*(z)=I_s$ and $\cs(\pP_j)\subseteq[-1,1]$;

\item[{\rm(3)}] $(\pP_{j+1}, \pP_j)$ has mutually compatible symmetry for all $0\le j\le J$;
%The symmetries of $\pP_{j}$ and $\pP_{j+1}$ are mutually compatible for all $0\le j\le J$;

\item[{\rm(4)}] $\pP_{0}=\pU_{\sym \pth_2}^*$ and $\pP_{J+1}=\mbox{diag}(\pU_{\sym \pth_1}, I_{s-r})$, where $\pU_{\sym \pth_1}$, $\pU_{\sym \pth_2}$ are products of a permutation matrix with a diagonal matrix of monomials, as defined in
\eqref{sym:reorganize};

\item[{\rm(5)}] $J\le \displaystyle \max_{1\le m \le r, 1\le n\le s} \lceil {|\cs([\pP]_{m,n})|}/{2}\rceil$, where $\lceil \cdot \rceil$ is the ceiling function.
\end{enumerate}
\end{theorem}

The representation in \eqref{cascade} (without symmetry) is often
called the cascade structure in the literature of engineering, see
\cite{Jiang1,Jiang2,V}. In the context of wavelet analysis, matrix
extension without symmetry has been discussed by Lawton, Lee and
Shen in their interesting paper \cite{LLS} and a simple algorithm
has been proposed there to derive a desired matrix $\pP_e$ from a
given row vector $\pP$ of Laurent polynomials without symmetry. In
electronic engineering, an algorithm using the cascade structure for
matrix extension without symmetry has been given in \cite{V} for
filter banks with perfect reconstruction property. The algorithms in
\cite{LLS,V} mainly deal with the special case that $\pP$ is a row
vector (that is, $r=1$ in our case) without symmetry and the
coefficient support of the derived matrix $\pP_e$ indeed can be
controlled by that of $\pP$. The algorithms in \cite{LLS,V} for the
special case $r=1$ can be employed to handle a general $r\times s$
matrix $\pP$ without symmetry, see \cite{LLS,V} for detail. However,
for the general case $r>1$, it is no longer clear whether the
coefficient support of the derived matrix $\pP_e$ obtained by the
algorithms in \cite{LLS,V} can still be  controlled by that of
$\pP$.

Several special cases of matrix extension with symmetry have been
considered in the literature. For $\F=\R$ and $r=1$, matrix
extension with symmetry has been considered in \cite{P}. For $r=1$,
matrix extension with symmetry has been studied in
\cite{Han:jfaa:2009} and a simple algorithm is given there. In the
context of wavelet analysis, several particular cases of matrix
extension with symmetry related to the construction of wavelets and
multiwavelets have been investigated in
\cite{Cui,Han:d4,Han:jfaa:2009,HJ:d4,Jiang1,Jiang2,P,Turca}.
However, for the general case of an $r\times s$ matrix, the
approaches on matrix extension with symmetry in
\cite{Han:jfaa:2009,P} for the particular case $r=1$ cannot be
employed to handle the general case. The algorithms in
\cite{Han:jfaa:2009,P} are very difficult to be generalized to the
general case $r>1$, partially due to the complicated relations of
the symmetry patterns between different rows of $\pP$. For the
general case of matrix extension with symmetry, it becomes much
harder to control the coefficient support of the derived matrix
$\pP_e$, comparing with the special case $r=1$. Extra effort is
needed in any algorithm of deriving $\pP_e$ so that its coefficient
support can be controlled by that of $\pP$.

The contributions of this paper lie in the following aspects.
Firstly, we satisfactorily solve the general matrix extension
problem with symmetry for any $r,s$ such that $1\le r\le s$. More
importantly, we obtain a complete representation of any $r\times s$
paraunitary matrix $\pP$ having compatible symmetry  with $1\le r\le
s$. This representation  leads to a step-by-step algorithm for
deriving a desired matrix $\pP_e$ from a given matrix $\pP$.
Secondly, we obtain an optimal result in the sense of
\eqref{supp:control} on controlling the coefficient support of the
desired matrix $\pP_e$ derived from a given matrix $\pP$ by our
algorithm.  This is of importance in both theory and application,
since short support of a filter or a multiwavelet is a highly
desirable property and short support usually means a fast algorithm
and simple implementation in practice. Thirdly, we introduce the
notion of compatibility of symmetry, which plays a critical role  in
the study of the general matrix extension problem with symmetry for
the multi-row  case ($r\ge1$). Fourthly, we provide a complete
analysis and a systematic construction algorithm for $\df$-band
symmetric filter banks and symmetric orthonormal multiwavelets.
Finally, most of the literature on the matrix extension problem only
consider Laurent polynomials with coefficients in the special field
$\C$ (\cite{LLS}) or $\R$ (\cite{CL,P}). In this paper, our setting
is under a  general field $\F$, which can be any subfield of $\C$
satisfying \eqref{F}.

The structure of this paper is as follows. In section~2, we shall
present a step-by-step algorithm which leads to constructive proofs
of Theorems~\ref{thm:main:1} and \ref{thm:main:2}. In section~3, we
shall discuss an application of our main results on matrix extension
with symmetry to the design of symmetric filter banks in electronic
engineering and to the construction of symmetric orthonormal
multiwavelets in wavelet analysis. Examples will be provided to
illustrate our algorithms. Finally, we shall prove
Theorems~\ref{thm:main:1} and \ref{thm:main:2} in section~4.

\section{An Algorithm for Matrix Extension with Symmetry}

In this section, we present a step-by-step algorithm on matrix
extension with symmetry to derive a desired matrix $\pP_e$ in
Theorem~\ref{thm:main:2} from a given matrix $\pP$. Our algorithm
has three steps: initialization, support reduction, and
finalization. The step of initialization reduces the symmetry
pattern of $\pP$ to a standard form. The step of support reduction
is the main body of the algorithm, producing a sequence of
elementary  matrices $\pA_1, \ldots, \pA_J$ that reduce the length
of the coefficient support of $\pP$ to $0$. The step of finalization
generates the desired matrix $\pP_e$ as in Theorem~\ref{thm:main:2}.
More precisely, our algorithm written in the form of
\emph{pseudo-code} for Theorem \ref{thm:main:2} is as follows:

\begin{algorithm} \label{alg:1} {
Input $\pP$ as in Theorem \ref{thm:main:2} with
$\sym\pP=(\sym\pth_1)^*\sym\pth_2$ for some $1\times r$ and $1\times
s$ row vectors $\pth_1$ and $\pth_2$ of Laurant polynomials with
symmetry.
%Output $\pP_e$ as in Theorem {\rm\ref{thm:main:2}}.
%\begin{itemize}
\\
\noindent \textbf{\rm 1. Initialization:} Let  $\pQ :=
\pU_{\sym\pth_1}^*\pP\pU_{\sym\pth_2}$. Then the symmetry pattern of
$\pQ$ is
\begin{equation}\label{Q:sym:standard}
\sym \pQ=[ {\bf1}_{ r_1},-{\bf1}_{ r_2 }, z{\bf1}_{ r_3 },
-z{\bf1}_{ r_4} ]^T[{\bf1}_{ s_1},-{\bf1}_{ s_2},  z^{-1}{\bf1}_{
s_3},-z^{-1}{\bf1}_{ s_4}],
\end{equation}
where all nonnegative integers $r_1,\ldots, r_4, s_1, \ldots, s_4$
are uniquely determined by $\sym \pP$.
\\
\noindent \textbf{\rm 2. Support Reduction:} Let
$\pP_0:=\pU_{\sym\pth_2}^*$ and $J:=1$.
\begin{tabbing}
\hspace*{0.16in}\=\hspace{3ex}\=\hspace{3ex}\=\hspace{3ex}\=\hspace{3ex}\=\hspace{3ex}\=\hspace{3ex}\kill
\> {\rm \texttt{while}} $(|\cs(\pQ)|>0)$ {\rm\texttt{do}}\qquad \texttt{\%\%} outer while loop\\
\> \> Let $\pQ_0:=\pQ$,  $[k_1,k_2]:=\cs(\pQ)$, and  $\pA_J:=I_s$.\\
\> \> {\rm\texttt{if}}  $k_2=-k_1$ {\rm\texttt{then}}\\
\> \> \> {\rm\texttt{for}} $j$ {\rm\texttt{from}} $1$ {\rm\texttt{to}} $r$ {\rm\texttt{do}}\\
\> \> \> \> Let $\pq:=[\pQ_0]_{j,:}$ and $\pp:=[\pQ]_{j,:}$, the $j$th rows of $\pQ_0$ and $\pQ$, respectively.\\
\> \> \> \> Let $[\ell_1,\ell_2]:=\cs(\pq)$, $\ell:=\ell_2-\ell_1$, and $\pB_j:=I_s$.\\
\> \> \> \> {\rm\texttt{if}} $\cs(\pq)=\cs(\pp)$ {\rm\texttt{and}} $\ell\ge2$ {\rm\texttt{and}} $(\ell_1=k_1$ {\rm\texttt{or}} $\ell_2=k_2)$ {\rm\texttt{then}}\\
\>  \> \> \> \>$\pB_j:=\pB_{\pq}$. $\pA_J:=\pA_J\pB_j$. $\pQ_0:=\pQ_0\pB_j$.\\
\>  \> \> \> {\rm\texttt{end if}}\\
\> \> \>{\rm\texttt{end for}}\\
\> \> \>$\pQ_0$ takes the form in \eqref{eq:Q0:form}. \\
\> \> \> Let $\pB_{(-k_2,k_2)}:=I_s$, $\pQ_1:=\pQ_0$, $j_1:=1$ and $j_2:=r_3+r_4+1$. \\
\> \> \>{\rm\texttt{while}} $j_1\le r_1+r_2$ {\rm\texttt{and}}  $j_2\le r$ {\rm\texttt{do}} \qquad \texttt{\%\%} inner while loop\\
\> \> \>\> Let $\pq_1:=[\pQ_1]_{j_1,:}$ and $\pq_2:=[\pQ_1]_{j_2,:}$. \\
\> \> \> \>  {\rm\texttt{if}} $\coeff(\pq_1,k_1)={\bf0}$ {\rm\texttt{then}} $j_1:=j_1+1$. {\rm\texttt{end if}}\\
\> \> \> \>  {\rm\texttt{if}} $\coeff(\pq_2,k_2)={\bf0}$ {\rm\texttt{then}} $j_2:=j_2+1$. {\rm\texttt{end if}}\\
\> \> \> \> {\rm\texttt{if}} $\coeff(\pq_1,k_1)\neq\bf0$ {\rm\texttt{and}} $\coeff(\pq_2,k_2)\neq \bf0$ {\rm\texttt{then}}\\
\> \> \> \>\>  $\pB_{(-k_2,k_2)}:=\pB_{(-k_2,k_2)}\pB_{(\pq_1,\pq_2)}$. $\pQ_1:=\pQ_1\pB_{(\pq_1,\pq_2)}$. $\pA_J:=\pA_J\pB_{(\pq_1,\pq_2)}$.\\
\> \> \> \>\>  $j_1:=j_1+1$. $j_2:=j_2+1$.\\
\> \> \> \> {\rm\texttt{end if}}\\
\> \> \>{\rm\texttt{end while}}\qquad \texttt{\%\%} end inner while loop\\
\> \> {\rm\texttt{end if}}\\
\> \>$\pQ_1$ takes the form in \eqref{eq:Q0:form} with either $\coeff(\pQ_1,-k)={\bf0}$ or $\coeff(\pQ_1,k)={\bf0}$.\\
\> \> Let $\pA_J:=\pA_J\pB_{\pQ_1}$ and $\pQ:=\pQ\pA_J$.\\
\> \> Then $\sym \pQ=[{\bf1}_{ r_1},-{\bf1}_{ r_2 },
 z{\bf1}_{ r_3 }, -z{\bf1}_{ r_4}
]^T[{\bf1}_{ s_1'},-{\bf1}_{ s_2'}, z^{-1}{\bf1}_{
s_3'},-z^{-1}{\bf1}_{ s_4'}]$.\\
\> \> Replace $s_1, \ldots, s_4$ by $s_1', \ldots, s_4'$, respectively. Let $\pP_J:=\pA_J^*$ and $J:=J+1$.\\
\> {\rm\texttt{end while}}\qquad \texttt{\%\%} end outer while loop
\end{tabbing}
\noindent \textbf{\rm 3. Finalization:}
%$\sym (\pQ)=[{\bf1}_{ r_1},-{\bf1}_{ r_2 }, z{\bf1}_{ r_3 },
%    -z{\bf1}_{ r_4},]^T[{\bf1}_{ s_1},-{\bf1}_{ s_2},  \frac{1}{z}{\bf1}_{s_3},-\frac{1}{z}{\bf1}_{ s_4}]$ and $|\cs(\pQ)|=0$. Then,
%Now \eqref{Q:sym:standard} holds and $|\cs(\pQ)|=0$. Hence
 $\pQ=\diag(F_1,F_2,F_3,F_4)$ for some $r_j\times s_j$ constant
matrices $F_j$ in $\F$, $j=1,\ldots,4$. Let
$U:=\diag(U_{F_1},U_{F_2},U_{F_3},U_{F_4})$ so that $\pQ
U=[I_r,{\bf0}]$. Define $\pP_{J}:=U^*$ and
$\pP_{J+1}:=\diag(\pU_{\sym\pth_1},I_{s-r})$.
\\
\noindent Output a desired matrix $\pP_e$
%:=\pP_{J}\pP_{J-1}\cdots\pP_1\pP_0$,
satisfying all the properties in Theorem~\ref{thm:main:2}.
%\end{itemize}
}\end{algorithm}

In the following subsections, we present detailed constructions of
the matrices $\pU_{\sym\pth}$, $\pB_{\pq}$, $\pB_{(\pq_1,\pq_2)}$,
$\pB_{\pQ_1}$, and $U_{F}$ appearing in Algorithm~\ref{alg:1}.

%
%The above algorithm produces a sequence of matrices
%$\pP_1,\cdots,\pP_J$. These matrices are paraunitary, compatible and
%reduce the support length of $\pP$ step by step from Lemma
%\ref{lemma:MatrixForvectorDegBy2} and Lemma
%\ref{lemma:MatrixforSpecialCase}. We shall also see in last section
%that each of $\pP_j$ is supported in $[-1,1]$. Together, we can
%prove the result of Theorem \ref{thm:main:1}. Before we present our
%proof, let us see the application of our results to the construction
%of symmetric filter banks and orthonormal multiwavelets.

\subsection{Initialization}
%Let $\pP$ be an $r\times s$ matrix with compatible symmetry, that is, \eqref{sym:comp} holds.
%for some $r \times 1$ matrix $\pp$ and $s\times 1$ matrix $\pq$ of Laurent polynomials with symmetry.
%To reduce the complexity, we first discuss how to reorganize the symmetry pattern of $\pP$ into a standard form.

Let $\pth$ be a $1\times n$ row vector of Laurent polynomials with
symmetry such that $\sym\pth=[\gep_1 z^{c_1}, \ldots, \gep_n
z^{c_n}]$ for some $\gep_1, \ldots, \gep_n\in \{-1,1\}$ and $c_1,
\ldots, c_n\in \Z$.  Then, the symmetry of any entry in the vector
$\pth \mbox{diag}(z^{-\lceil c_1/2\rceil}, \ldots, z^{-\lceil
c_n/2\rceil})$ belongs to $\{ \pm 1, \pm z^{-1}\}$. Thus, there is a
permutation matrix $E_\pth$ to regroup these four types of
symmetries together so that
\begin{equation}\label{sym:reorganize}
%\begin{small}
\sym(\pth \pU_{\sym\pth})=[\mathbf{1}_{n_1}, -\mathbf{1}_{n_2},
z^{-1} \mathbf{1}_{n_3}, -z^{-1}\mathbf{1}_{n_4}],
%\end{small}
\end{equation}
where $\pU_{\sym\pth}:= \mbox{diag}(z^{-\lceil c_1/2\rceil}, \ldots,
z^{-\lceil c_n/2\rceil})E_\pth$, $\mathbf{1}_m$ denotes the $1\times
m$ row vector $[1,\ldots,1]$,  and  $n_1, \ldots, n_4$ are
nonnegative integers uniquely determined by $\sym\pth$. Since $\pP$
satisfies \eqref{sym:comp}, it is easy to see that $\pQ:=\pU_{\sym
\pth_1}^*\pP \pU_{\sym \pth_2}$ has the symmetry pattern as in
\eqref{Q:sym:standard}. Note that $\pU_{\sym\pth_1}$ and
$\pU_{\sym\pth_2}$ do not increase the length of the coefficient
support of $\pP$.
%
%\begin{equation}\label{P:sym:reorganize}
%\sym (\pU_{\sym \pth_1}^*\pP \pU_{\sym \pth_2})= [\mathbf{1}_{r_1},
%-\mathbf{1}_{ r_2 }, z \mathbf{1}_{ r_3 },
%    -z\mathbf{1}_{ r_4}, ]^T
%[\mathbf{1}_{s_1},-\mathbf{1}_{ s_2},  z^{-1} \mathbf{1}_{
%s_3},-z^{-1}\mathbf{1}_{ s_4}],
%\end{equation}
%
%where all integers $r_1, \ldots, r_4, s_1, \ldots, s_4$ are completely determined by $\sym \pP$.

\subsection{Support Reduction} Denote $\pQ:=\pU_{\sym \pth_1}^*\pP \pU_{\sym \pth_2}$ as in Algorithm~\ref{alg:1}. The outer \texttt{while} loop in the step of support reduction produces
a sequence of elementary paraunitary matrices $\pA_1,\ldots, \pA_J$
that reduce the length of the coefficient support of $\pQ$ gradually
to $0$. The construction of each $\pA_j$ has three parts:
$\{\pB_1,\ldots,\pB_r\}$, $\pB_{(-k,k)}$, and $\pB_{\pQ_1}$. The
first part $\{\pB_1,\ldots,\pB_r\}$ (see the \texttt{for} loop) is
constructed recursively for each of the $r$ rows  of $\pQ$ so that
$\pQ_0:=\pQ\pB_1\cdots\pB_r$ has a special form as in
\eqref{eq:Q0:form}. If both $\coeff(\pQ_0,-k)\neq{\bf0}$ and
$\coeff(\pQ_0,k)\neq{\bf0}$, then the second part $\pB_{(-k,k)}$
(see the inner \texttt{while} loop) is further constructed so that
$\pQ_1:=\pQ_0\pB_{(-k,k)}$ takes the form in \eqref{eq:Q0:form} with
at least one of $\coeff(\pQ_1,-k)$ and $\coeff(\pQ_1,k)$ being
${\bf0}$. $\pB_{\pQ_1}$ is  constructed to handle the case that
$\cs(\pQ_1)=[-k,k-1]$ or $\cs(\pQ_1)=[-k+1,k]$ so that
$\cs(\pQ_1\pB_{\pQ_1})\subseteq[-k+1,k-1]$.

Let $\pq$ denote an arbitrary row of $\pQ$ with $|\cs(\pq)|\ge 2$.
We first explain how to construct $\pB_{\pq}$ for a given row $\pq$
such that $\pB_{\pq}$ reduces the  length of the coefficient support
of $\pq$ by $2$ and keeps its symmetry pattern. Note that in the
\texttt{for} loop, $\pB_j$ is simply $\pB_{\pq}$ with $\pq$ being
the current $j$th row of $\pQ\pB_0\cdots\pB_{j-1}$, where
$\pB_0:=I_s$.
%$\pB_0:=I_s$.
%Note that in the
%\texttt{for} loop, $\pB_j$ is simply $\pB_\pq$ with $\pq$ being the
%current $j$th row of $\pQ\pB_0\cdots\pB_{j-1}$ and  with
%$\pB_0:=I_s$.

By \eqref{Q:sym:standard}, we have $\sym \pq=\gep
z^{c}[\mathbf{1}_{s_1},-\mathbf{1}_{ s_2}, z^{-1} \mathbf{1}_{
s_3},-z^{-1}\mathbf{1}_{ s_4}]$ for some $\gep\in \{-1,1\}$ and
$c\in \{0,1\}$. For $\gep=-1$, there is a permutation matrix
$E_{\gep}$ such that $\sym(\pq E_{\gep})=
z^{c}[\mathbf{1}_{s_2},-\mathbf{1}_{ s_1}, z^{-1} \mathbf{1}_{
s_4},-z^{-1}\mathbf{1}_{ s_3}]$. For $\gep=1$, we let
$E_{\gep}:=I_{s}$. Then, $\pq E_{\gep}$ must take the form in either
\eqref{eq:type1} or \eqref{eq:type2} with $\vf_1\neq {\bf0}$ as
follows:
%
%\begin{equation}\label{eq:type1}
\begin{align}
\begin{split}
\pq E_{\gep}=&[ \vf_1, -\vf_2,  \vg_1, -\vg_2]z^{\ell_1} +
[\vf_3,-\vf_4,\vg_3,-\vg_4 ]z^{\ell_1+1}
+\sum_{\ell=\ell_1+2}^{\ell_2-2}\coeff(\pq E_{\gep},\ell)z^\ell
\\&+ [\vf_3,\vf_4,\vg_1,\vg_2]z^{\ell_2-1} +
[\vf_1,\vf_2,\textbf{0},{\bf0}]z^{\ell_2};
\end{split}\label{eq:type1}\\
\begin{split}
 \pq
E_{\gep}=&[{\bf0},{\bf0},\vf_1,- \vf_2]z^{\ell_1} + [ \vg_1,-
\vg_2,\vf_3,- \vf_4]z^{\ell_1+1}
+\sum_{\ell=\ell_1+2}^{\ell_2-2}\coeff(\pq E_{\gep},\ell)z^\ell\\&+
[ \vg_3, \vg_4, \vf_3,\vf_4]z^{\ell_2-1}+ [ \vg_1, \vg_2,  \vf_1,
\vf_2]z^{\ell_2}.
\end{split}\label{eq:type2}
\end{align}
%\end{equation}
%
If $\pq E_{ \gep}$ takes the form in \eqref{eq:type2}, we further
construct a permutation matrix $E_{\pq}$ such that $[ \vg_1, \vg_2,
\vf_1, \vf_2]E_{\pq}=[\vf_1, \vf_2,\vg_1,\vg_2]$ and  define
$\pU_{\pq,\gep}:=E_{\gep} E_{\pq}\diag(I_{s-s_{\vg}},z^{-1}
I_{s_{\vg}})$, where $s_{\vg}$ is the size of the row vector
$[\vg_1,\vg_2]$. Then $\pq \pU_{\pq,\gep}$ takes the form in
\eqref{eq:type1}.  For $\pq E_{\gep}$ of form \eqref{eq:type1}, we
simply let $\pU_{\pq,\gep}:=E_{\gep}$. In this way,
$\pq_0:=\pq\pU_{\pq,\gep}$ always takes the form in \eqref{eq:type1}
with $\vf_1\neq{\bf0}$.

Note that $\pU_{\pq,\gep}\pU_{\pq,\gep}^*=I_s$ and
$\|\vf_1\|=\|\vf_2\|$ if $\pq_0\pq_0^*=1$, where
$\|\vf\|:=\sqrt{\vf\vf^*}$. Now we construct an $s\times s$
paraunitary matrix $\pB_{\pq_0}$ to reduce the coefficient support
of $\pq_0$  as in \eqref{eq:type1} from $[\ell_1,\ell_2]$ to
$[\ell_1+1,\ell_2-1]$ as follows:
\begin{equation}\label{eq:MatrixForvectorDegBy2}
%\begin{small}
\pB_{\pq_0}^*:=\frac{1}{c}\left[
 \begin{array}{c|c|c|c}
 \vf_1(z+\frac{{c_0}}{c_{\vf_1}}+\frac1z)& \vf_2(z-\frac1z) & \vg_1(1+\frac1z) & \vg_2(1-\frac1z) \\
c F_1&  {\bf 0} &{\bf 0} &{\bf 0}\\
\hline
\noalign{\vspace{-0.1in}}&&&\\
-\vf_1(z-\frac1z) & -\vf_2(z-\frac{{c_0}}{c_{\vf_1}}+\frac1z) & -\vg_1(1-\frac1z) & -\vg_2(1+\frac1z) \\
{\bf 0}& c F_2  &{\bf 0} &{\bf 0}\\
\hline
\noalign{\vspace{-0.1in}}&&&\\
\frac{c_{\vg_1}}{c_{\vf_1}}\vf_1(1+z) & -\frac{c_{\vg_1}}{c_{\vf_1}}\vf_2(1-z) & c_{\vg_1'}\vg_1' & {\bf 0}\\
{\bf 0}&  {\bf 0} & c G_1&{\bf 0}\\
\hline
\noalign{\vspace{-0.1in}}&&&\\
\frac{c_{\vg_2}}{c_{\vf_1}}\vf_1(1-z) & -\frac{c_{\vg_2}}{c_{\vf_1}}\vf_2(1+z) & {\bf 0} & c_{\vg_2'} \vg_2'\\
{\bf 0}&  {\bf 0} &{\bf 0} &c G_2\\
\end{array}
\right],
%\end{small}
\end{equation}
where $c_{\vf_1}:=\|\vf_1\|$,  $c_{\vg_1}:=\|\vg_1\|$, $
c_{\vg_2}:=\|\vg_2\|$, $c_0:=\coeff(\pq_0,\ell_1+1)\coeff(\pq_0^*,-\ell_2)/c_{\vf_1}$,%{(\vf_3 \vf_1^*-\vf_4\vf_2^*)}/{c_{\vf_1}}$,
%$c:=\sqrt{4c_{\vf}^2+2(c_{\vg_1}^2+c_{\vg_2}^2)+|c_0|^2}$.
\begin{equation}
\begin{small}
\begin{aligned}
&c_{\vg_1'}:=
\begin{cases} \frac{-2c_{\vf_1}-\overline{c_0}}{c_{\vg_1}} &\text{if $\vg_1\neq{\bf0}$;}\\ c &\text{otherwise,}\end{cases}
\qquad
c_{\vg_2'}:=\begin{cases} \frac{2c_{\vf_1}-\overline{c_0}}{c_{\vg_2}} &\text{if $\vg_2\neq {\bf0}$;}\\
c &\text{otherwise,}
\end{cases}\\
&c:=(4c_{\vf_1}^2+2c_{\vg_1}^2+2c_{\vg_2}^2+|c_0|^2)^{1/2},
\end{aligned}
\end{small}
\end{equation}
and $[\frac{\vf_j^*}{\|\vf_j\|},F_j^*]=U_{\vf_j}$,
$[\vg_j'^*,G_j^*]=U_{\vg_j}$ for $j=1,2$ are unitary constant
extension matrices in $\F$ for vectors $\vf_j,\vg_j$ in $\F$,
respectively (see section~4 for a concrete construction of  such
unitary matrices $U_{\vf_j}$ and $U_{\vg_j}$). Here, the role of a
unitary constant matrix $U_{\vf}$ in $\F$ is to reduce the number of
nonzero entries in $\vf$ such that $\vf
U_{\vf}=[\|\vf\|,0,\ldots,0]$. The operations for the emptyset
$\emptyset$ are defined by $\|\emptyset\|=\emptyset$,
$\emptyset+A=A$ and $\emptyset\cdot A=\emptyset$ for any object $A$.

Define
$\pB_{\pq}:=\pU_{\pq,\varepsilon}\pB_{\pq_0}\pU_{\pq,\varepsilon}^*$.
Then $\pB_{\pq}$ is paraunitary. Due to the particular form of
$\pB_{\pq_0}$ as in \eqref{eq:MatrixForvectorDegBy2}, direct
computations yield the following very important properties of the
paraunitary matrix $\pB_\pq$:
\begin{itemize}
\item[\rm{(P1)}]  $\sym\pB_{\pq}=[\mathbf{1}_{s_1},-\mathbf{1}_{ s_2},
z\mathbf{1}_{ s_3},-z\mathbf{1}_{
s_4}]^T[\mathbf{1}_{s_1},-\mathbf{1}_{ s_2}, z^{-1} \mathbf{1}_{
s_3},-z^{-1}\mathbf{1}_{ s_4}]$, $\cs({\pB}_\pq)=[-1,1]$, and
$\cs(\pq\pB_{\pq})=[\ell_1+1,\ell_2-1]$. That is, $\pB_\pq$ has
compatible symmetry with coefficient support on $[-1,1]$ and
$\pB_{\pq}$ reduces the length of the coefficient support of $\pq$
exactly by $2$. Moreover, $\sym(\pq\pB_{\pq})=\sym\pq$.

\item[\rm{(P2)}] if $(\pp, \pq^*)$ has mutually compatible symmetry and
$\pp\pq^*=0$, then $\sym(\pp\pB_{\pq})=\sym(\pp)$ and $\cs(\pp
\pB_\pq)\subseteq \cs(\pp)$. That is, $\pB_\pq$ keeps the symmetry
pattern of $\pp$ and does not increase the  length of the
coefficient support of  $\pp$.
\end{itemize}

Next, let us explain the construction of $\pB_{(-k,k)}$. For
$\cs(\pQ)=[-k,k]$  with $k\ge 1$, $\pQ$ is of the form as follows:
\begin{equation}\label{eq:P:Standardform}
\begin{scriptsize}
\begin{aligned}
\pQ
&= \left[
  \begin{array}{cccc}
      F_{11} & -F_{21} & G_{31} & -G_{41} \\
    -F_{12} & F_{22} &  -G_{32} & G_{42} \\
    \hline
   {\bf0} & {\bf0}&   F_{31} & -F_{41}  \\
   {\bf0} & {\bf0}&   -F_{32} & F_{42}\\
  \end{array}
\right]z^{-k}+\left[
  \begin{array}{cccc}
    F_{51} & -F_{61} &  G_{71} & -G_{81} \\
    -F_{52} & F_{61} & -G_{72} & G_{82}  \\
        \hline
     G_{11} & -G_{21}&F_{71} & -F_{81} \\
    -G_{12} & G_{22} &-F_{72} & F_{82}   \\
  \end{array}
\right]z^{-k+1} \\&+
 \sum_{n=2-k}^{k-2}\coeff(\pQ,n)
 +\left[
  \begin{array}{cccc}
    F_{51} & F_{61} & G_{31} & G_{41} \\
    F_{52} & F_{61} & G_{32} & G_{42} \\
        \hline
     G_{51} & G_{61}&F_{71} & F_{81} \\
    G_{52} & G_{62} &F_{72} & F_{82}   \\
  \end{array}
\right]z^{k-1}+\left[
  \begin{array}{cccc}
     F_{11} & F_{21} & {\bf0} & {\bf0} \\
    F_{12} & F_{22} &  {\bf0} & {\bf0}\\
        \hline
     G_{11} & G_{21}&F_{31} & F_{41} \\
    G_{12} & G_{22} &F_{32} & F_{42}   \\
  \end{array}
\right]z^{k}
\end{aligned}
\end{scriptsize}
\end{equation}
with all $F_{jk}$'s and $G_{jk}$'s being constant matrices in $\F$
and $F_{11},F_{22},F_{31},F_{42}$ being  of size $r_1\times s_1$,
$r_2\times s_2$, $r_3\times s_3$, $r_4\times s_4$, respectively. Due
to Property (P1) and (P2) of $\pB_{\pq}$, the \texttt{for} loop in
Algorithm \ref{alg:1} reduces $\pQ$ in \eqref{eq:P:Standardform} to
$\pQ_0:=\pQ\pB_1\cdots\pB_r$ as follows:
\begin{equation}
\label{eq:Q0:form}
\begin{scriptsize}
\begin{aligned}
 \left[
  \begin{array}{cccc}
   {\bf0} & {\bf0}& \wt G_{31} & -\wt G_{41} \\
  {\bf0} & {\bf0}&  -\wt G_{32} & \wt G_{42} \\
    \hline
    \noalign{\vspace{-0.09in}}\\
   {\bf0} & {\bf0}&   {\bf0} & {\bf0}  \\
   {\bf0} & {\bf0}&  {\bf0} & {\bf0}\\
  \end{array}
\right]z^{-k}
%+\left[
%  \begin{array}{cccc}
%    F_{51} & -F_{61} &  G_{71} & -G_{81} \\
%    -F_{52} & F_{61} & -G_{72} & G_{82}  \\
%        \hline
%     G_{11} & G_{21}&F_{71} & -F_{81} \\
%    G_{12} & G_{22} &-F_{72} & F_{82}   \\
%  \end{array}
%\right]z^{-k+1} \\&
+\cdots
%\sum_{n=1-k}^{k-1}\coeff(\pQ_0,n)
% +\left[
%  \begin{array}{cccc}
%    F_{51} & F_{61} & G_{31} & G_{41} \\
%    F_{52} & F_{61} & G_{32} & G_{42} \\
%        \hline
%     G_{51} & G_{61}&F_{71} & F_{81} \\
%    G_{52} & G_{62} &F_{72} & F_{82}   \\
%  \end{array}
%\right]z^{k-1}
+\left[
  \begin{array}{cccc}
    {\bf0} & {\bf0}& {\bf0} & {\bf0} \\
   {\bf0} & {\bf0} &  {\bf0} & {\bf0}\\
        \hline
    \noalign{\vspace{-0.09in}}\\
     \wt G_{11} & \wt G_{21}&{\bf0} & {\bf0} \\
    \wt G_{12} & \wt G_{22} &{\bf0} & {\bf0}  \\
  \end{array}
\right]z^{k}.
\end{aligned}
\end{scriptsize}
\end{equation}
If either $\coeff(\pQ_0,-k)={\bf0}$ or $\coeff(\pQ_0,k)={\bf0}$,
then the inner \texttt{while} loop does nothing and
$\pB_{(-k,k)}=I_s$. If both $\coeff(\pQ_0,-k)\neq{\bf0}$ and
$\coeff(\pQ_0,k)\neq{\bf0}$, then $\pB_{(-k,k)}$ is constructed
recursively from pairs $(\pq_1,\pq_2)$ with $\pq_1,\pq_2$ being two
rows of $\pQ_0$ satisfying $\coeff(\pq_1,-k)\neq {\bf0}$ and
$\coeff(\pq_2,k)\neq {\bf0}$. The construction of
$\pB_{(\pq_1,\pq_2)}$ with respect to such a pair $(\pq_1,\pq_2)$ in
the inner \texttt{while} loop is as follows.
%\begin{itemize}
%\item[{\rm(a)}]

Similar to the discussion before \eqref{eq:type1}, there is a
permutation matrix $E_{(\pq_1,\pq_2)}$ such that
$\wt\pq_1:=\pq_1E_{(\pq_1,\pq_2)}$ and
$\wt\pq_2:=\pq_2E_{(\pq_1,\pq_2)}$ take the following form:
\begin{equation}
\label{eq:pqform}
\begin{scriptsize}
\begin{aligned}
&\left[
  \begin{array}{c}
    \wt \pq_1 \\
\wt \pq_2 \\
  \end{array}
\right]=   \left[
  \begin{array}{cccc}
      {\bf0}& {\bf0} & \wt g_{3} &  -\wt g_{4} \\
        \hline
    \noalign{\vspace{-0.09in}}\\
   {\bf0} & {\bf0}&   {\bf0} & {\bf0}  \\
  \end{array}
\right]z^{-k}+\left[
  \begin{array}{cccc}
   \wt  f_{5} & -\wt f_{6} & \wt  g_{7} & -\wt g_{8} \\
        \hline
    \noalign{\vspace{-0.09in}}\\
     \gep\wt g_{1} & -\gep\wt g_{2}&\gep\wt f_{7} & -\gep\wt f_{8} \\
  \end{array}
\right]z^{-k+1} \\ &\qquad+%\cdots
 \sum_{n=2-k}^{k-2}\coeff(\left[
  \begin{array}{c}
    \wt \pq_1 \\
\wt \pq_2 \\
  \end{array}
\right],n)
 +\left[
  \begin{array}{cccc}
   \wt  f_{5} & \wt f_{6} & \wt g_{3} &\wt g_{4} \\
        \hline
    \noalign{\vspace{-0.09in}}\\
     \wt g_{5} &\wt g_{6}&\wt f_{7} & \wt f_{8} \\
  \end{array}
\right]z^{k-1}+\left[
  \begin{array}{cccc}
     {\bf0} & {\bf0} & {\bf0} & {\bf0}  \\
        \hline
    \noalign{\vspace{-0.09in}}\\
     \wt g_{1} &\wt  g_{2}&{\bf0} & {\bf0} \\
  \end{array}
\right]z^{k},
\end{aligned}
\end{scriptsize}
\end{equation}
where $\gep\in\{-1,1\}$ and all $\wt\vg_j$'s are  nonzero row
vectors. Note that $\|\wt\vg_1\|=\|\wt\vg_2\|=:c_{\wt\vg_1}$ and
$\|\wt\vg_3\|=\|\wt\vg_4\|=:c_{\wt\vg_3}$. Construct an $s\times s$
paraunitary matrix $\pB_{(\wt\pq_1,\wt\pq_2)}$ as follows:

\begin{equation}\label{eq:MatrixForvectorDegBy2SpecialCase}
\begin{small}
\pB_{(\wt\pq_1,\wt\pq_2)}^*:=\frac{1}{c}\left[
 \begin{array}{c|c|c|c}
 \frac{{c_0}}{c_{\wt\vg_1}}\wt \vg_1& {\bf0} & \wt\vg_3(1+\frac1z) & \wt\vg_4(1-\frac1z) \\
\noalign{\vspace{-0.1in}}&&&\\
c\wt G_1&  {\bf 0} &{\bf 0} &{\bf 0}\\
\hline
\noalign{\vspace{-0.1in}}&&&\\
{\bf 0} & \frac{{c_0}}{c_{\wt\vg_1}}\wt\vg_2 & -\wt\vg_3(1-\frac1z) & -\wt\vg_4(1+\frac1z) \\
\noalign{\vspace{-0.1in}}&&&\\
{\bf 0}&c\wt G_2  &{\bf 0} &{\bf 0}\\
\hline
\noalign{\vspace{-0.1in}}&&&\\
\frac{c_{\wt\vg_3}}{c_{\wt\vg_1}}\wt\vg_1(1+z) & -\frac{c_{\wt\vg_3}}{c_{\wt\vg_1}}\wt\vg_2(1-z) & -\frac{\overline{c_0}}{c_{\wt\vg_3}}\wt\vg_3 & {\bf 0}\\
\noalign{\vspace{-0.1in}}&&&\\
{\bf 0}&  {\bf 0} &c\wt G_3&{\bf 0}\\
\hline
\noalign{\vspace{-0.1in}}&&&\\
\frac{c_{\wt\vg_3}}{c_{\wt\vg_1}}\wt\vg_1(1-z) & -\frac{c_{\wt\vg_3}}{c_{\wt\vg_1}}\wt\vg_2(1+z) & {\bf 0} & -\frac{\overline{c_0}}{c_{\wt\vg_3}}\wt \vg_4\\
\noalign{\vspace{-0.1in}}&&&\\
{\bf 0}&  {\bf 0} &{\bf 0} &c\wt G_4\\
\end{array}
\right],
\end{small}
\end{equation}
where $c_0:=\coeff(\wt\pq_1,-k+1)\coeff(\wt\pq_2^*,-k)/c_{\wt\vg_1}$, %(\wt\vf_5\wt \vg_1^*-\wt\vf_6\wt\vg_2^*)/c_{\wt\vg_1}$,
$c:=(|c_0|^2+4c_{\wt\vg_3}^2)^{1/2}$, and
$[\frac{\wt\vg_j^*}{\|\wt\vg_j\|},\wt G_j^*]=U_{\wt\vg_j}$  are
unitary constant extension matrices in $\F$ for vectors $\wt\vg_j$
in $\F$, $j=1,\ldots,4$, respectively. Let
$\pB_{(\pq_1,\pq_2)}:=E_{(\pq_1,\pq_2)}\pB_{(\wt\pq_1,\wt\pq_2)}E_{(\pq_1,\pq_2)}^T$.
Similar to Property (P1) and (P2) of $\pB_{\pq}$, we have the
following very important properties of $\pB_{(\pq_1,\pq_2)}$:
\begin{itemize}
\item[\rm{(P3)}]  $\sym\pB_{(\pq_1,\pq_2)}=[\mathbf{1}_{s_1},-\mathbf{1}_{ s_2}, z \mathbf{1}_{
s_3},-z\mathbf{1}_{ s_4}]^T[\mathbf{1}_{s_1},-\mathbf{1}_{ s_2},
z^{-1} \mathbf{1}_{ s_3},-z^{-1}\mathbf{1}_{ s_4}]$, the coefficient
support of ${\pB}_{(\pq_1,\pq_2)}$ is on $[-1,1]$,
$\cs(\pq_1\pB_{(\pq_1,\pq_2)})\subseteq[-k+1,k-1]$ and
$\cs(\pq_2\pB_{(\pq_1,\pq_2)})\subseteq[-k+1,k-1]$. That is,
$\pB_{(\pq_1,\pq_2)}$ has compatible symmetry with coefficient
support on $[-1,1]$ and $\pB_{(\pq_1,\pq_2)}$ reduces the length of
both the coefficient supports of $\pq_1$ and $\pq_2$  by $2$.
Moreover, $\sym(\pq_1\pB_{(\pq_1,\pq_2)})=\sym\pq_1$ and
$\sym(\pq_2\pB_{(\pq_1,\pq_2)})=\sym\pq_2$.

\item[\rm{(P4)}] if both $(\pp, \pq_1^*)$ and $(\pp,\pq_2^*)$ have mutually compatible symmetry and
$\pp\pq_1^*=\pp\pq_2^*=0$, then
$\sym(\pp\pB_{(\pq_1,\pq_2)})=\sym\pp$ and $\cs(\pp
\pB_{(\pq_1,\pq_2)})\subseteq \cs(\pp)$. That is,
$\pB_{(\pq_1,\pq_2)}$ keeps the symmetry pattern of $\pp$ and does
not increase the  length of the coefficient support of  $\pp$.
\end{itemize}

Now, due to the Property (P3) and (P4) of $\pB_{(\pq_1,\pq_2)}$,
$\pB_{(-k,k)}$ constructed in the inner \texttt{while} loop reduces
$\pQ_0$ of the form in \eqref{eq:Q0:form} with both
$\coeff(\pQ_0,-k)\neq{\bf0}$ and $\coeff(\pQ_0,k)\neq{\bf0}$, to
$\pQ_1:=\pQ_0\pB_{(-k,k)}$ of the form in \eqref{eq:Q0:form} with
either
  $\coeff(\pQ_1,-k)=\coeff(\pQ_1,k)={\bf0}$ (for
this case, simply let $\pB_{\pQ_1}:=I_s$) or one of
$\coeff(\pQ_1,-k)$ and $\coeff(\pQ_1,k)$ is nonzero. For the latter
case, $\pB_{\pQ_1}:=\diag(U_1\pW_1,I_{s_3+s_4})E$ with matrices
$U_1,\pW_1$ constructed with respect to $\coeff(\pQ_1,k)\neq {\bf0}$
or $\pB_{\pQ_1}:=\diag(I_{s_1+s_2},U_3\pW_3)E$ with $U_3,\pW_3$
constructed with respect to $\coeff(\pQ_1,-k)\neq {\bf0}$, where $E$
is a permutation matrix. $\pB_{\pQ_1}$ is constructed so that
$\cs(\pQ_1\pB_{\pQ_1})\subseteq[-k+1,k-1]$. Let $\pQ_1$ take form in
\eqref{eq:Q0:form}. The matrices $U_1,\pW_1$ or $U_3, \pW_3$, and
$E$ are constructed as follows.
%\begin{itemize}
%\item[{\rm(a)}]

Let $U_1:=\diag(U_{\wt G_1},U_{\wt G_2})$ and  $U_3:=\diag(U_{\wt
G_3},U_{\wt G_4})$ with
\begin{equation}\label{eq:wtG}
\begin{small}
\wt G_1:=\left[
          \begin{array}{c}
            \wt  G_{11} \\
            \wt  G_{12} \\
          \end{array}
        \right],\,
\wt G_2:=\left[
          \begin{array}{c}
            \wt  G_{21} \\
            \wt  G_{22} \\
          \end{array}
        \right],\,
\wt G_3:=\left[
          \begin{array}{c}
            \wt  G_{31} \\
            \wt  G_{32} \\
          \end{array}
        \right],\,
\wt G_4:=\left[
          \begin{array}{c}
            \wt  G_{41} \\
            \wt  G_{42} \\
          \end{array}
        \right].
\end{small}
\end{equation} Here, for a nonzero matrix $G$ with rank $m$,  $U_G$ is a
unitary matrix such that $GU_{G}=[R,{\bf0}]$ for some matrix $R$ of
rank $m$. For $G={\bf0}, U_{G}:=I$ and for $G=\emptyset,
U_{G}:=\emptyset$. When $G_1G_1^*=G_2G_2^*$, $U_{G_1}$ and $U_{G_2}$
can be constructed such that $G_1U_{G_1}=[R,{\bf0}]$ and
$G_2U_{G_2}=[R,{\bf0}]$ (see section~4 for more detail).

%Let $\wt G_1:=[\wt G_{11};\wt G_{12}]$, $G_2:=[\wt G_{21};\wt
%G_{22}]$, $G_3:=[\wt G_{31}; \wt G_{32}]$ and $G_4:=[\wt G_{41}; \wt
%G_{42}]$.
%Here a block matrix $A=[B;C]$ from $B$ and $C$ is a matrix with
%first row $B$ and second row $C$. Let $m_1$ be the rank of $G_1$ and
%$m_3$ be the
% rank of $G_3$. Due to the  orthogonality of $\pQ$,  we
%have $G_1G_1^*=G_2G_2^*$, $G_3G_3^*=G_4G_4^*$. Hence, there exist
%two lower triangle matrices $R_1, R_3$ such that $G_1 U_{G_1}
%=[R_1,{\bf0}], G_2 U_{G_2}=[R_1,{\bf0}]$ and $G_3
%U_{G_3}=[R_3,{\bf0}]$, $G_4U_4=[R_3,{\bf0}]$ with $R_1$ of size
%$(r_3+r_4)\times m_1$ and $R_3$ of size $(r_1+r_2)\times m_3$. Let
%$U:=\diag(U_{G_1},U_{G_2},U_{G_3},U_{G_4})$.

%\item[{\rm(b)}]

Let $m_1$, $m_3$ be the ranks of $\wt G_1$,   $\wt G_3$,
respectively ($m_1=0$ when $\coeff(\pQ_1,k)={\bf0}$ and $m_3=0$ when
$\coeff(\pQ_1,-k)={\bf0}$). Note that $\wt G_1\wt G_1^*=\wt G_2\wt
G_2^*$ or $\wt G_3\wt G_3^*=\wt G_4\wt G_4^*$ due to
$\pQ_1\pQ_1^*=I_r$. The matrices $\pW_1,\pW_3$ are then constructed
by:
\begin{equation}
\label{eq:W1W3}
\begin{small}
\begin{aligned}
\pW_1:=\left[
           \begin{array}{c|c|c|c}
             \pU_1 & & \pU_2&  \\
             \hline
             & I_{s_1-{m_1}} & &    \\
              \hline
             \pU_2 & & \pU_1 &\\
              \hline
            & & & I_{s_2-m_1}\\
           \end{array}
         \right],
\pW_3:=\left[
           \begin{array}{c|c|c|c}
            \pU_3 & & \pU_4& \\
             \hline
              & I_{s_3-{m_3}} & &    \\
              \hline
              \pU_4 & & \pU_3 & \\
              \hline
              & & & I_{s_4-m_3} \\
                \end{array}
\right],
\end{aligned}
\end{small}
\end{equation}
where $\pU_1(z)=-\pU_2(-z):=\frac{1+z^{-1}}{2}I_{m_1}$ and
$\pU_3(z)=\pU_4(-z):=\frac{1+z}{2}I_{m_3}$.

%\item[{\rm(c)}]

Let $\pW_{\pQ_1}:=\diag(U_1\pW_1,I_{s_3+s_4})$ for the case that
$\coeff(\pQ_1,k)\neq{\bf0}$ or
$\pW_{\pQ_1}:=\diag(I_{s_1+s_2},U_3\pW_3)$ for the case that
$\coeff(\pQ_1,-k)\neq{\bf0}$. Then $\pW_{\pQ_1}$ is paraunitary. By
the symmetry pattern and orthogonality of $\pQ_1$, $\pW_{\pQ_1}$
reduces the coefficient support of $\pQ_1$ to $[-k+1,k-1]$, i.e.,
$\cs(\pQ_1\pW_{\pQ_1})=[-k+1,k-1]$. Moreover, $\pW_{\pQ_1}$ changes
the symmetry pattern of $\pQ_1$ such that
 $\sym(\pQ_1 \pW_{\pQ_1})=[
    {\bf1}_{ r_1},
    -{\bf1}_{ r_2 },
 z{\bf1}_{ r_3 },
    -z{\bf1}_{ r_4}
]^T\sym\pth_1$ with
\[
\sym \pth_1=[ z^{-1}{\bf1}_{m_1},{\bf1}_{s_1-m_1}, -z^{-1}{\bf1}_{
m_1},-{\bf1}_{s_2-m_1 },{\bf1}_{m_3}, z^{-1}{\bf1}_{s_3-m_3
},-{\bf1}_{m_3}, -z^{-1}{\bf1}_{s_4-m_3}].
\]
 $E$ is then the permutation matrix
such that
\[
\sym (\pQ_1 \pW_{\pQ_1})E=[
    {\bf1}_{ r_1},
    -{\bf1}_{ r_2 },
 z{\bf1}_{ r_3 },
    -z{\bf1}_{ r_4},
]^T\sym\pth,
\] with
$\sym \pth=[ {\bf1}_{s_1-m_1+m_3},,-{\bf1}_{s_2-m_1+m_3},
z^{-1}{\bf1}_{s_3-m_3+m_1},
-z^{-1}{\bf1}_{s_4-m_3+m_1}]=(\sym\pth_1) E$.
%%\end{itemize}

\section{Application to Filter Banks and Orthonormal
Multiwavelets with Symmetry}

In this section, we shall discuss the application of our results on
matrix extension with symmetry to $\df$-band symmetric paraunitary
filter banks in electronic engineering and to orthonormal
multiwavelets with symmetry in wavelet analysis. In order to do so,
let us introduce some definitions first.

We say that $\df$ is {\it a dilation factor} if $\df$ is an integer
with $|\df|>1$. Throughout this section, $\df$ denotes a dilation
factor. For simplicity of presentation, we further assume that $\df$
is positive, while multiwavelets and filter banks with a negative
dilation factor can be handled similarly by a slight modification of
the statements in this paper.

Let $\F$ be a subfield of $\C$ such that \eqref{F} holds. A low-pass
filter $\ta_0: \Z \mapsto \F^{\mphi\times \mphi}$ with multiplicity
$\mphi$ is a finitely supported sequence of $\mphi\times \mphi$
matrices on $\Z$. The \emph{symbol} of the filter $\ta_0$ is defined
to be $\pa_0(z):=\sum_{k\in \Z} \ta_0(k) z^k$, which is a matrix of
Laurent polynomials with coefficients in $\F$. Moreover, the
\emph{$\df$-band subsymbols} of $\ta_0$ are defined by $\pa_{0;
\gamma}(z):=\sqrt{\df} \sum_{k\in\Z} \ta_0(\gamma+\df k) z^k$,
$\gamma\in \Z$. We say that $\pa_0$ (or $\ta_0$) is {\it a
$\df$-band orthogonal filter} if
\begin{equation}\label{mask:orth}
%[\pa_{0;0}(z), \ldots, \pa_{0; \df-1}(z)][\pa_{0;0}(z), \ldots, \pa_{0; \df-1}(z)]^*=
\sum_{\gamma=0}^{\df-1} \pa_{0;\gamma}(z) \pa_{0; \gamma}^*(z)=I_\mphi, \qquad z\in \C \bs \{0\}.
\end{equation}
To design an orthogonal filter bank with the perfect reconstruction property, one has to design high-pass filters $\ta_1, \ldots, \ta_{\df-1}: \Z \mapsto \F^{\mphi\times \mphi}$ such that the polyphase matrix
\begin{equation}\label{polyphase}
\PP(z)=\left[ \begin{matrix} \pa_{0;0}(z) &\cdots &\pa_{0; \df-1}(z)\\
\pa_{1;0}(z) &\cdots &\pa_{1; \df-1}(z)\\
\vdots &\vdots &\vdots\\
\pa_{\df-1;0}(z) &\cdots &\pa_{\df-1; \df-1}(z)
\end{matrix}\right]
\end{equation}
is paraunitary, that is, $\PP(z) \PP^*(z)=I_{\df\mphi}$, where each
$\pa_{m;\gamma}$ is a subsymbol of $\pa_m$ for
$m,\gamma=0,\ldots,\df-1$, respectively. Symmetry of the filters in
a filter bank is a very much desirable property in many
applications. We say that the low-pass filter $\pa_0$ (or $\ta_0$)
has symmetry if
\begin{equation}\label{mask:sym}
\pa_0(z)=\mbox{diag}(\gep_1 z^{\df c_1}, \ldots, \gep_{\mphi} z^{\df
c_{\mphi}}) \pa_0(1/z) \mbox{diag}(\gep_1 z^{-c_1}, \ldots,
\gep_{\mphi} z^{-c_{\mphi}})
\end{equation}
for some $\gep_1, \ldots, \gep_{\mphi}\in \{-1,1\}$ and $c_1,
\ldots, c_{\mphi}\in \R$ such that $\df c_\ell-c_j\in\Z$ for all
$\ell,j=1,\ldots,r$. To design a symmetric filter bank with the
perfect reconstruction property, from a given $\df$-band orthogonal
low-pass filter $\ta_0$, one has to construct high-pass filters
$\ta_1, \ldots, \ta_{\df-1}: \Z \mapsto \F^{\mphi\times \mphi}$ such
that all of them have  symmetry that is compatible with  the
symmetry of $\pa_0$ in \eqref{mask:sym} and the polyphase matrix
$\PP$ in \eqref{polyphase} is paraunitary.

For $f\in L_1(\R)$, the Fourier transform used in this paper is
defined to be $\hat f(\xi):=\int_\R f(x) e^{-i x\xi} dx$ and can be
naturally extended to $L_2(\R)$ functions. For a $\df$-band
orthogonal low-pass filter $\pa_0$, we assume that there exists {\it
an orthogonal $\df$-refinable function vector} $\phi=[\phi_1,
\ldots, \phi_\mphi]^T$ associated with the low-pass filter $\pa_0$,
with compactly supported functions $\phi_1, \ldots, \phi_\mphi$ in
$L_2(\R)$, such that
\begin{equation}\label{F:refeq}
\hat \phi(\df \xi)=\pa_0(e^{-i\xi}) \hat \phi(\xi), \qquad \xi\in \R \qquad \mbox{with}\quad \|\hat \phi(0)\|=1,
\end{equation}
and
\begin{equation}\label{orth:reffunc}
\la \phi(\cdot -k), \phi\ra:=\int_\R \phi(x-k) \ol{\phi(x)}^T dx=\gd(k) I_\mphi, \qquad k\in \Z,
\end{equation}
where $\gd$ denotes the {\it Dirac sequence} such that $\gd(0)=1$
and $\gd(k)=0$ for all $k\ne 0$. Define  multiwavelet function
vectors $\psi^m=[\psi^m_1,\ldots,\psi^m_r]^T$ associated with the
high-pass filters $\pa_m$, $m=1,\ldots,\df-1$, by
\begin{equation}\label{wavelet}
\wh{\psi^m}(\df\xi):=\pa_m(e^{-i\xi}) \wh{\phi}(\xi), \qquad \xi\in
\R, \; m=1, \ldots, \df-1.
\end{equation}
It is well known that $\{\psi^1, \ldots, \psi^{\df-1}\}$ generates
an orthonormal multiwavelet basis in $L_2(\R)$; that is, $\{ \df
^{j/2}\psi^m_{\ell}(\df^j\cdot-k)\; : \; j, k\in \Z; m=1, \ldots,
\df-1; \ell=1,\ldots,r\}$ is an orthonormal basis of $L_2(\R)$, for
example, see \cite{Daub:book,HKZ,HZ2009,Shen} and references
therein.

If $\pa_0$ has symmetry as in \eqref{mask:sym} and if $1$ is a
simple eigenvalue of $\pa_0(1)$, then it is well known that the
$\df$-refinable function vector $\phi$ in \eqref{F:refeq} associated
with the low-pass filter $\pa_0$ has the following symmetry:
\begin{equation}\label{phi:sym}
\phi_1(c_1-\cdot)=\gep_1 \phi_1,\quad \phi_2(c_2-\cdot)=\gep_2
\phi_2,\quad \ldots, \quad \phi_\mphi(c_\mphi-\cdot)=\gep_\mphi
\phi_\mphi.
\end{equation}

Under the symmetry condition in \eqref{mask:sym}, to apply Theorem
\ref{thm:main:1}, we first show that there exists  a suitable
paraunitary matrix $\pU$ acting on
$\pP_{\pa_0}:=[\pa_{0;0},\ldots,\pa_{0;\df-1}]$ so that
$\pP_{\pa_0}\pU$ has compatible symmetry. Note that $\pP_{\pa_0}$
itself may not have any symmetry.

\begin{lemma}\label{lemma:symPa}
Let $\pP_{\pa_0}:=[\pa_{0;0},\ldots,\pa_{0;\df-1}]$, where
$\pa_{0;0}, \ldots, \pa_{0; \df-1}$ are $\df$-band subsymbols of a
$\df$-band orthogonal filter $\pa_0$ satisfying \eqref{mask:sym}.
Then there exists a $\df r\times \df r$ paraunitary matrix $\pU$
such that $\pP_{\pa_0}\pU$ has compatible symmetry.
\end{lemma}

\begin{proof}
From \eqref{mask:sym}, we deduce that
% have
%$[a_0(k)]_{\ell,j}=\varepsilon_\ell\varepsilon_j[a_0(\df
%c_\ell-c_j-k)]_{\ell,j}$ for all $k\in \Z$. Then
%
\begin{equation}\label{eq:polyPhaseSymCondition}
[\pa_{0;\gamma}(z)]_{\ell,j}=\varepsilon_\ell\varepsilon_jz^{R_{\ell,j}^\gamma}[\pa_{0;{Q_{\ell,j}^\gamma}}(z^{-1})]_{\ell,j},\,
\gamma=0,\ldots,\df-1; \ell,j=1,\ldots,r,
\end{equation}
%
%\[
%\begin{aligned}
% [\pa_{0;\gamma}(z)]_{\ell,j} =%{\sqrt{\df}}\sum_{k\in\Z}[a_0({m+\df
%k})]_{\ell,j}z^k
%\\&={\sqrt{\df}}\sum_{k\in\Z}\varepsilon_{\ell,j}[a_0({\df c_\ell-c_j-m-\df k})]_{\ell,j}z^k
%\\&={\sqrt{\df}}\sum_{k\in\Z}\varepsilon_{\ell,j}[a_0({\df (-k+R_{\ell,j}^m)+Q_{\ell,j}^m})]_{\ell,j}z^k
%\\&=\varepsilon_{\ell,j}z^{R_{\ell,j}^m}{\sqrt{d}}\sum_{k\in\Z}[a_0({\df k+Q_{\ell,j}^m})]_{\ell,j}z^{-k}
%\\&=
%\varepsilon_\ell\varepsilon_jz^{R_{\ell,j}^\gamma}[\pa_{0;{Q_{\ell,j}^\gamma}}(z)]_{\ell,j},
%\end{aligned}
%\]
where $\gamma,Q_{\ell,j}^\gamma\in \Gamma:=\{0,\ldots,\df-1\}$ and
$R_{\ell,j}^\gamma$, $Q_{\ell,j}^\gamma$ are uniquely determined by
\begin{equation}\label{RQ}
\df c_\ell-c_j-\gamma=\df R_{\ell,j}^\gamma+Q_{\ell,j}^\gamma \quad
\mbox{with} \quad R_{\ell,j}^\gamma\in \Z, \;
Q_{\ell,j}^\gamma\in\Gamma.
\end{equation}
Since $\df c_\ell-c_j\in\Z$ for all $\ell, j=1,\ldots, r$, we have
$c_\ell-c_j\in\Z$ for all $\ell,j=1,\ldots,r$ and therefore,
$Q_{\ell,j}^\gamma$ is independent of $\ell$. Consequently, by
\eqref{eq:polyPhaseSymCondition}, for every $1\le j\le r$, the $j$th
column of the matrix $\pa_{0;\gamma}$ is a flipped version of the
$j$th column of the matrix $\pa_{0;{Q_{\ell,j}^\gamma}}$.
%Using some simple linear combinations, it is very easy to find some invertible elementary transform acting on these two columns to obtain two vectors of Laurent polynomials with symmetry. In order to control the coefficient support, some shifts on the polyphases might be needed.
Let $\kappa_{j,\gamma}\in\Z$ be an integer such that
$|\cs([\pa_{0;\gamma}]_{:,j}
+z^{\kappa_{j,\gamma}}[\pa_{0;{Q_{\ell,j}^\gamma}}]_{:,j})|$
%(or $|\cs(z^{k_{j,\gamma}}[\pa_{0;\gamma}]_{:,j} +[\pa_{0;{Q_{\ell,j}^\gamma}}]_{:,j}$)
is as small as possible. Define
$\pP:=[\pb_{0;0},\ldots,\pb_{0;\df-1}]$ as follows:
\begin{equation}\label{eq:symmetrization}
[\pb_{0;\gamma}]_{:,j}:= \begin{cases}
      [\pa_{0;\gamma}]_{:,j}, & \hbox{$\gamma=Q_{\ell,j}^\gamma;$} \\
       \frac{1}{\sqrt{2}}([\pa_{0;\gamma}]_{:,j}+ z^{\kappa_{j,\gamma}}[\pa_{0;{Q_{\ell,j}^\gamma}}]_{:,j}), & \hbox{$\gamma<Q_{\ell,j}^\gamma$;}\\
       \frac{1}{\sqrt{2}}([\pa_{0;\gamma}]_{:,j}-z^{\kappa_{j,\gamma}}[\pa_{0;{Q_{\ell,j}^\gamma}}]_{:,j}), & \hbox{$\gamma>Q_{\ell,j}^\gamma$,}
       \end{cases}
\end{equation}
where $[\pa_{0;\gamma}]_{:,j}$ denotes the $j$th column of
$\pa_{0;\gamma}$. Let $\pU$ denote the unique transform matrix
corresponding to \eqref{eq:symmetrization} such that
$\pP:=[\pb_{0;0},\ldots,\pb_{0;\df-1}]=[\pa_{0;0},\ldots,\pa_{0;\df-1}]
\pU$. It is evident that $\pU$ is paraunitary and
$\pP=\pP_{\pa_0}\pU$. We now show that $\pP$ has compatible
symmetry. Indeed, by \eqref{eq:polyPhaseSymCondition} and
\eqref{eq:symmetrization},
\begin{equation}\label{eq:symPatternPb}
[\sym \pb_{0;\gamma}]_{\ell,j}=
\sgn(Q_{\ell,j}^\gamma-\gamma)\gep_\ell\gep_jz^{R_{\ell,j}^\gamma+\kappa_{j,\gamma}},
\end{equation}
where $\sgn(x)=1$ for $x\ge 0$ and $\sgn(x)=-1$ for $x<0$.
By \eqref{RQ} and noting that $Q_{\ell,j}^\gamma$ is independent of
$\ell$, we have
\[
\frac{[\sym \pb_{0;\gamma}]_{\ell,j}}{[\sym \pb_{0;\gamma}]_{n,j}}
%=\frac{\pm\varepsilon_{m,j}z^{R_{m,j}^\gamma+k_{j,\gamma}}}{\pm
%\varepsilon_{n,j}z^{R_{n,j}^\gamma+k_{j,\gamma}}}
=\gep_\ell\gep_nz^{R_{\ell,j}^\gamma-R_{n,j}^\gamma}=
\gep_\ell\gep_nz^{c_\ell-c_n},
\]
for all $1\le \ell,n\le r$,
% Thus, we have $[\sym
%\pP]_{\ell,:}=\gep_\ell\gep_n z^{c_{\ell}-c_{n}}[\sym \pP]_{n,:}$,
which is equivalent to saying that $\pP$ has compatible symmetry.
\end{proof}

Now, for a $\df$-band orthogonal low-pass filter $\pa_0$ satisfying
\eqref{mask:sym}, we have the following algorithm to construct
high-pass filters $\pa_1,\ldots,\pa_{\df-1}$ such that they form a
symmetric paraunitary filter bank with the perfect reconstruction
property.

\begin{algorithm}
\label{alg:2} Input an orthogonal $\df$-band filter $\pa_0$ with
symmetry in \eqref{mask:sym}.
\begin{itemize}
\item[{\rm(1)}] Construct  $\pU$ as in \eqref{eq:symmetrization} such
that $\pP:=\pP_{\pa_0}\pU$ has compatible symmetry:
$\sym\pP=[\gep_1z^{k_1},\ldots,\gep_r z^{k_r}]^T\sym\pth$ for some
$k_1,\ldots,k_r\in\Z$ and some $1\times \df r$ row vector $\pth$ of
Laurent polynomials with symmetry.

\item[{\rm(2)}] Derive $\pP_e$ with all the properties as in Theorem~\ref{thm:main:1} from $\pP$ by
Algorithm~\ref{alg:1}.

\item[{\rm(3)}] Let $\PP:=\pP_e\pU^*=:(\pa_{m;\gamma})_{0\le m,\gamma\le \df-1}$ as  in \eqref{polyphase}. Define high-pass filters
\begin{equation}
\label{highpass:def}
\pa_m(z):=\frac{1}{\sqrt{\df}}
\sum_{\gamma=0}^{\df-1}\pa_{m;\gamma}(z^\df)z^\gamma, \qquad m=1, \ldots, \df-1.
\end{equation}
\end{itemize}
Output a symmetric filter bank $\{ \pa_0, \pa_1, \ldots,
\pa_{\df-1}\}$ with the perfect reconstruction property, i.e. $\PP$
in \eqref{polyphase} is paraunitary and all filters $\pa_m$, $m=1,
\ldots, \df-1$, have symmetry:
\begin{equation}\label{highpass:sym}
\pa_{m}(z)=\diag(\gep^m_1z^{\df
c^m_1},\ldots,\gep^m_rz^{\df
c^m_r})\pa_m(1/z)\diag(\gep_1z^{-c_1},\ldots,\gep_{r}z^{-c_r}),
\end{equation}
where $c^m_\ell:=(k^m_{\ell}-k_\ell)+c_{\ell}\in\R$ and all
$\varepsilon^m_{\ell}\in\{-1,1\}$, $k^m_{\ell}\in\Z$, for
$\ell=1,\ldots,r$ and  $m=1,\ldots,\df-1$, are determined by the
symmetry pattern of $\pP_e$ as follows:
\begin{equation}
\label{sym:Pe} [\gep_1 z^{k_1},\ldots,\gep_r z^{k_r}, \gep^1_1
z^{k^1_1},\ldots,\gep^1_r
z^{k^1_r},\ldots,z^{k^{\df-1}_1},\ldots,\gep^{\df-1}_r
z^{k^{\df-1}_r}]^T\sym\pth:=\sym\pP_e.
\end{equation}

%$\sym\pP_e:=[\sym\pth_0,\sym\pth_1,\ldots,\sym\pth_{\df-1}]^T(\sym\pth)$
%with $\pth$ being an $1\times (\df r)$ row vector of Laurent
%polynomials with symmetry and $\pth_0,\ldots,\pth_{\df-1}$ all being
%$1\times r$ row vector of Laurent polynomials with symmetry given by
%%\[
%%\sym\pth_0=[\gep_1 z^{k_1},\ldots,\gep_r z^{k_r}],\;
%%\sym\pth_1=[\gep^1_1 z^{k^1_1},\ldots,\gep^1_r z^{k^1_r}], \;
%%\cdots,\; \sym\pth_{\df-1}=[\gep^{\df-1}_1
%%z^{k^{\df-1}_1},\ldots,\gep^{\df-1}_r z^{k^{\df-1}_r}].
%%\]
%\begin{equation}
%\label{sym:Pe}
%\begin{aligned}
%\sym\pth_0&=[\gep_1 z^{k_1},\ldots,\gep_r z^{k_r}],\\
%\sym\pth_1&=[\gep^1_1 z^{k^1_1},\ldots,\gep^1_r z^{k^1_r}],
%\\
%&\;\,\vdots\\
%\sym\pth_{\df-1}&=[\gep^{\df-1}_1
%z^{k^{\df-1}_1},\ldots,\gep^{\df-1}_r z^{k^{\df-1}_r}].
%\end{aligned}
%\end{equation}

\end{algorithm}

\begin{proof} Rewrite $\pP_e=(\pb_{m;\gamma})_{0\le m,\gamma \le\df-1}$ as a $\df\times \df$ block matrix
with $r\times r$ blocks $\pb_{m;\gamma}$. Since $\pP_e$ has
compatible symmetry as in \eqref{sym:Pe}, we have $[\sym
\pb_{m;\gamma}]_{\ell,:}=\varepsilon^m_\ell\varepsilon_\ell
z^{k^m_{\ell}-k_{\ell}}[\sym \pb_{0;\gamma}]_{\ell,:}$ for
$\ell=1,\ldots, r$ and $m=1,\ldots,\df-1$. By
\eqref{eq:symPatternPb}, we have
\begin{equation}
\label{eq:symCompatibleB}
[\sym\pb_{m;\gamma}]_{\ell,j}=\sgn(Q_{\ell,j}^\gamma-\gamma)\varepsilon^m_\ell\varepsilon_jz^{R_{\ell,j}^\gamma+\kappa_{j,\gamma}+k^m_\ell-k_\ell},
\qquad  \ell, j=1,\ldots,r.
\end{equation}
By \eqref{eq:symCompatibleB} and the definition of $\pU^*$ in \eqref{eq:symmetrization}, we deduce that
\begin{equation}\label{eq:compatibleCondition}
[\pa_{m;\gamma}]_{\ell,j} =
\varepsilon^m_\ell\varepsilon_jz^{R_{\ell,j}^\gamma+k^m_\ell-k_\ell}[\pa_{m;{Q_{\ell,j}^\gamma}}(z^{-1})]_{\ell,j}.
\end{equation}
This implies that
$[\sym\pa_m]_{\ell,j}=\varepsilon^m_\ell\varepsilon_j
z^{\df(k^m_\ell-k_\ell+c_\ell)-c_j}$, which is equivalent to
\eqref{highpass:sym} with $c^m_{\ell}:=k^m_{\ell}-k_\ell+c_\ell$ for
$m=1,\ldots,\df-1$ and $\ell = 1,\ldots,r$.
\end{proof}

Since the high-pass filters $\pa_1,\ldots,\pa_{\df-1}$ satisfy
\eqref{highpass:sym}, it is easy to verify that each
$\psi^m=[\psi^m_1, \ldots, \psi^m_r]^T$ defined in \eqref{wavelet}
also has the following symmetry:
\begin{equation}\label{sym:psi}
\psi^m_{1}(c^m_{1}-\cdot)=\varepsilon^m_{1}\psi^m_{1},\quad
\psi^m_{2}(c^m_{2}-\cdot)=\varepsilon^m_{2}\psi^m_{2},\quad
\ldots,\quad\psi^m_{r}(c^m_{r}-\cdot)=\varepsilon^m_{r}\psi^m_{r}.
\end{equation}

 In the following, let us present several
examples to demonstrate our results and illustrate our algorithms.
\begin{example}\label{ex:extGHM}
{\rm Let $\df=2$ and $r=2$. A $2$-band orthogonal low-pass filter
$\pa_0$  with multiplicity $2$ in \cite{GHM} is given by
\[
\pa_0(z)= \frac{1}{40}\left[
\begin{array}{cc}12(1+z^{-1})& 16\sqrt {2}z^{-1}\\
-\sqrt {2}({z}^{2}-9z-9+z^{-1})&-2(3z-10+3z^{-1})
\end{array}
\right].
\]
The low-pass filter $\pa_0$ satisfies \eqref{mask:sym} with $c_1=-1,
c_2=0$ and $\varepsilon_1=\varepsilon_2=1$. Using Lemma
\ref{lemma:symPa}, we obtain $\pP_{\pa_0}:=[\pa_{0;0},\pa_{0;1}]$
and $\pU$  as follows:
\[
\begin{small}
\begin{aligned}
\pP_{\pa_0}=\frac{1}{20}\left[ \begin {array}{cc|cc} 6\sqrt
{2}&0&\frac{6\sqrt {2}}{z}&\frac{16}{z}\\
\noalign{\vspace{-0.05in}}& \\
 9-z&10\sqrt {2}&9-\frac1z&-3\sqrt
{2}(1+\frac1z)\end {array} \right]
\end{aligned},\,
\pU= \frac{1}{\sqrt{2}}\left[
       \begin{array}{cccc}
         1& 0 &1 & 0 \\
         0 & \sqrt{2} & 0 & 0 \\
         z & 0 & -z & 0 \\
         0 & 0 & 0 & \sqrt{2}z \\
       \end{array}
     \right].
\end{small}
\]
Then $\pP:=\pP_{\pa_0}\pU$ satisfies
$\sym\pP=[1,z]^T[1,z^{-1},-1,1]$ and is given by
\[
\pP=\frac{\sqrt{2}}{20}
\left[
\begin{array}{cccc}
6\sqrt{2}&0&0&8\sqrt {2}\\
4(1+z)&10&5(1-z)&-3(1+z)
\end{array}
\right].
\]
Applying Algorithm~\ref{alg:1}, we obtain a desired paraunitary
matrix $\pP_e$ as follows:
\[
\pP_e=\frac{\sqrt{2}}{20} \left[
\begin{array}{cccc}
6\sqrt {2}&0&0&8\sqrt {2}\\
4(1+z)&10&5(1-z)&-3(1+z)\\
4(1+z)&-10&5(1-z)&-3(1+z)\\
4\sqrt{2}(1-z) &0&5\sqrt {2} (z+1) &3\sqrt {2}(z-1)\\
\end{array}
\right].
\]
We have $\sym\pP_e=[1,z,z,-z]^T[1,z^{-1},-1,1]$ and
$\cs([\pP_e]_{:,j})\subseteq \cs([\pP]_{:,j})$ for all $1\le j\le
4$. Now, from the polyphase matrix
$\PP:=\pP_e\pU^*=:(\pa_{m;\gamma})_{0\le m, \gamma\le 1}$, we derive
a high-pass filter $\pa_1$ as follows:
\[
\pa_1(z)=\frac{1}{40} \left[
\begin{array}{cc}
 -\sqrt{2}({z}^{2}-9z-9+z^{-1})&-2(3z+10+3z^{-1})\\
2({z}^{2}-9z+9-z^{-1})&6\sqrt {2}(z-z^{-1})
\end{array}
\right].
\]
Then the high-pass filter $\pa_1$ satisfies \eqref{highpass:sym}
with $c^1_1=c^1_2=0$ and $\varepsilon^1_1=1, \varepsilon^1_2=-1$.
%See Figure~3.1 for the
%graphs  of the $2$-refinable function vector $\phi$ associated with
%the low-pass filter $\pa_0$ and the multiwavelet function vector
%$\psi$ associated with the high-pass filter $\pa_1$.
%\begin{figure}[th] \label{fig:extGHM}
%\centerline{\scalebox{1.0}{
%\hbox{\epsfig{file=./ext0PhiPsi.eps,width=3.9in,height=1.6in} }}}
%\begin{caption}
%\,The graphs of the $2$-refinable function vector
%$\phi=[\phi_1,\phi_2]^T$ associated with the low-pass filter $\pa_0$
%(top row) and the multiwavelet function vector
%$\psi=[\psi_1,\psi_2]^T$ associated with the high-pass filter
%$\pa_1$ (bottom row) in Example \ref{ex:extGHM}.
%%$\phi_1=\phi_1(-1-\cdot), \phi_2=\phi_2(-\cdot)$ and
%%$\psi_1=\psi_1(-\cdot), \psi_2=-\psi_2(-\cdot)$.
%\end{caption}
%\end{figure}
}
\end{example}

\begin{example}\label{ex:RefinableInterpolantd3r2Sym01}
{\rm
 Let $\df=3$ and $r=2$. A $3$-band orthogonal low-pass filter
 $\pa_0$  with multiplicity $2$ in
 \cite{HZ2009} is given by:
\[
\pa_0(z)=\frac{1}{540}\left[
           \begin{array}{cc}
             a_{11}(z)+a_{11}(z^{-1}) & a_{12}(z)+z^{-1} a_{12}(z^{-1}) \\
              a_{21}(z)+z^3a_{21}(z^{-1}) &  a_{22}(z)+z^2 a_{22}(z^{-1}) \\
           \end{array}
         \right],
\]where
\[
\begin{small}
\begin{aligned}
a_{11}(z)&=90+(55-5\sqrt{41})z-(8+2\sqrt{41})z^2+(
7\sqrt{41}-47)z^{4};\\
a_{12}(z)&=145+5\sqrt {41}+ ( 1-\sqrt {41} )
{z}^{2}+( 34-4\sqrt{41} ) {z}^{3};\\
a_{21}(z)&= ( 111+9\sqrt {41}) {z}^{2}+( 69-9\sqrt {41}) {z}^{4};
\\
a_{22}(z)&=90z+( 63-3\sqrt {41} ) {z}^{2}+(3\sqrt {41}-63) {z}^{3}.
\end{aligned}
\end{small}
\]
The low-pass filter $\pa_0$ satisfies \eqref{mask:sym} with $c_1=0,
c_2=1$ and $\varepsilon_1=\varepsilon_2=1$. From
$\pP_{\pa_0}:=[\pa_{0;0},\pa_{0;1},\pa_{0;2}]$,  the matrix $\pU$
constructed by Lemma \ref{lemma:symPa} is given by
\[
\pU:=
\frac{1}{\sqrt{2}}
\left[
\begin{array}{cccccc}
\sqrt{2}&0&0&0&0&0\\
0&1&0&0&0&1\\
0&0&1&0 &1&0\\
0&0&0&\sqrt{2}&0&0\\
0&0&z&0&-z&0\\
0&z&0&0&0&-z
\end{array}
\right].
\]
Let
\[
\begin{small}
\begin{aligned}
c_0&=11-\sqrt{41};&
t_{12}&=5(7-\sqrt{41});&
c_{12}&=10(29+\sqrt{41});&
t_{13}&=-5c_0;\\
t_{16}&=3c_0;&
t_{15}&=3(3\sqrt{41}-13);&
t_{25}&=6(7+3\sqrt{41});&
t_{26}&=6(21-\sqrt{41});\\
t_{53}&=400\sqrt{6}/c_0;&
t_{55}&=12\sqrt{6}(\sqrt {41}-1);&
t_{56}&=6\sqrt {6}(4+\sqrt {41} );&
c_{66}&=3\sqrt{6}(3+7\sqrt{41}).
\end{aligned}
\end{small}
\]
Then $\pP:=\pP_{\pa_0}\pU$ satisfies
$\sym\pP=[1,z]^T[1,1,1,z^{-1},-1,-1]$ and is given by
\[
\pP=\frac{\sqrt{6}}{1080}
\left[
\begin {array}{cccccc}
180\sqrt{2}& b_{12}(z)& b_{13}(z)&0&
t_{15}(z-z^{-1})&t_{16}(z-z^{-1})
\\% \noalign{\medskip}
0&0&180(1+z)&180\sqrt {2}& t_{25}(1-z)& t_{26}(1-z)
\end {array}
\right],
\]where $b_{12}(z)=t_{12}(z+z^{-1})+c_{12}$ and
$b_{13}(z)=t_{13}(z-2+z^{-1})$. Applying Algorithm~\ref{alg:1}, we
obtain a desired paraunitary matrix $\pP_e$ as follows:
\[
\begin{small}
\pP_e=
\frac{\sqrt{6}}{1080}
\left[
\begin {array}{cccccc}
180\sqrt{2}& b_{12}(z)& b_{13}(z)&0&
t_{15}(z-\tfrac1z)&t_{16}(z-\tfrac1z)
\\% \noalign{\medskip}
0&0&180(1+z)&180\sqrt {2}& t_{25}(1-z)& t_{26}(1-z)\\
\hline
360&-\frac{b_{12}(z)}{\sqrt{2}}&-\frac{b_{13}(z)}{\sqrt{2}}&0 &\frac{t_{15}}{\sqrt{2}}(\tfrac1z-z)&\frac{t_{16}}{\sqrt{2}}(\tfrac1z-z)\\
%\noalign{\medskip}
0&0&90\sqrt {2}(1+z)&-360&\frac{t_{25}}{\sqrt{2}}(1-z)&\frac{t_{26}}{\sqrt{2}}(1-z)\\
\hline
0&\sqrt{6}t_{13}(1-z) &t_{53}(1-z) &0&t_{55}(1+z)&t_{56}(1+z)
\\
0&\frac{\sqrt{6}t_{12}}{2}(\tfrac1z-z)&\frac{\sqrt{6}t_{13}}{2}(\tfrac1z-z)&0&b_{65}(z)&b_{66}(z)
\end {array} \right],
\end{small}
\]
where  $b_{65}(z)= -\sqrt{6}(5t_{15}(z+z^{-1})+3c_{12})/10$ and
$b_{66}(z)=-\sqrt{6}t_{16}(z+z^{-1})/2+c_{66}$.
%\[
%\begin{aligned}
%t_{32}&= 5(\sqrt {41}-7)/\sqrt {2};&
%
%c_{32}&=-5\sqrt{2}(29+\sqrt{41});&
%
%t_{33}&=5(11-\sqrt{41})/\sqrt{2};\\
%
%c_{33}&=5\sqrt{2}(\sqrt {41}-11);&
%
%t_{35}&=3(13-3\sqrt {41})/\sqrt {2};&
%
%t_{36}&=3(\sqrt {41}-11)/\sqrt {2};\\
%
%t_{45}&=3\sqrt {2} ( 7+3\sqrt {41} );&
%
%t_{46}&=3\sqrt {2} ( 21-\sqrt {41});&
%
%t_{52}&=5\sqrt {6} ( 11-\sqrt {41} );\\
%
%t_{53}=5\sqrt{6} ( 11+\sqrt {41} );
%%
%t_{55}=12\sqrt{6}(\sqrt {41}-1);
%%
%t_{56}=6\sqrt {6}(4+\sqrt {41} );
%
%t_{62}&=5\sqrt {6}(\sqrt {41}-7)/2;&
%
%t_{63}&=5\sqrt {6 }(11-\sqrt {41})/2;&
%
%t_{65}&=\sqrt{6}(39-9\sqrt{41})/2;\\
%
%c_{65}&=-\sqrt {6}(87+3\sqrt{41});&
%
%t_{66}&=3\sqrt{6}(\sqrt{41}-11)/2;&
%
%c_{66}=\sqrt{6}(9+21\sqrt{41}).
%\end{aligned}
%\]
Note that $\sym\pP_e=[1,z,1,z,-z,-1]^T[1,1,1,z^{-1},-1,-1]$ and the
coefficient support of $\pP_e$ satisfies
$\cs([\pP_e]_{:,j})\subseteq\cs([\pP]_{:,j})$ for all $1\le j\le 6$.
From the polyphase matrix $\PP:=\pP_e\pU^*=:(\pa_{m;\gamma})_{0\le
m, \gamma\le 2}$, we derive two high-pass filters $\pa_1, \pa_2$ as
follows:
\[
\begin{aligned}
\pa_1(z)&=\frac{\sqrt{2}}{1080}\left[
           \begin{array}{cc}
             a^1_{11}(z)+a^1_{11}(z^{-1}) & a^1_{12}(z)+z^{-1} a^1_{12}(z^{-1}) \\
             \noalign{\vspace{-0.1in}}\\
              a^1_{21}(z)+z^3a^1_{21}(z^{-1}) &  a^1_{22}(z)+z^2 a^1_{22}(z^{-1}) \\
           \end{array}
         \right],\\
\pa_2(z)&=\frac{\sqrt{6}}{1080}\left[
           \begin{array}{cc}
             a^2_{11}(z)-z^3a^2_{11}(z^{-1}) & a^2_{12}(z)-z^2 a^2_{12}(z^{-1}) \\
             \noalign{\vspace{-0.1in}}\\
              a^2_{21}(z)-a^2_{21}(z^{-1}) &  a^2_{22}(z)-z^{-1} a^2_{22}(z^{-1}) \\
           \end{array}
         \right],
\end{aligned}
\]where
\[
\begin{small}
\begin{aligned}
a^1_{11}(z)&=( 47-7\sqrt {41} ) {z}^{4}+ 2( 4+\sqrt {41}
 ) {z}^{2}+ 5(\sqrt {41} -11) z+180;\\
a^1_{12}(z)&=2( 2\sqrt {41} -17) {z}^{3}+ ( \sqrt {41}-1 )
{z}^{2}-5(29+\sqrt {41})
;\\
\end{aligned}
\end{small}
\]
\[
\begin{small}
\begin{aligned}
a^1_{21}(z)&= 3(37+3\sqrt {41} )z+3(23-3\sqrt {41})z^{-1};
\\
a^1_{22}(z)&=-180z+3(21-\sqrt {41})-3(21-\sqrt {41})z^{-1};\\
%
%
%mask2
a^2_{11}(z)&=(43+17\sqrt{41})z+(67-7\sqrt {41})z^{-1};\\
a^2_{12}(z)&=11\sqrt {41}-31-(79+\sqrt {41})z^{-1};\qquad\qquad\qquad\\
a^2_{21}(z)&= ( 47-7\sqrt {41} ) {z}^{4}+ 2( 4+\sqrt {41} ) {z}^{2}-
3(29+\sqrt {41} ) z ;\qquad\quad
\\
a^2_{22}(z)&=2( 2\sqrt {41}-17 ) {z}^{3}+ ( \sqrt {41}-1 )
{z}^{2}+3(3+7\sqrt {41}).
\end{aligned}
\end{small}
\]
Then the high-pass filters $\pa_1$, $\pa_2$ satisfy
\eqref{highpass:sym} with $c^1_1=0, c^1_2=1$,
$\varepsilon^1_1=\varepsilon^1_2=1$ and $c^2_1=1, c^2_2=0$,
$\varepsilon^2_1=\varepsilon^2_2=-1$.
}
\end{example}

As demonstrated by the following example, our Algorithm \ref{alg:2}
also applies to low-pass filters with symmetry patterns other than
those in \eqref{mask:sym}.

\begin{example}\label{ex:RefinableInterpolantd3r2SymQuarter}
{\rm Let $\df=3$ and $r=2$. A $3$-band orthogonal low-pass filter
$\pa_0$ with multiplicity $2$ in \cite{HKZ} is given by
\[
\pa_0(z)=\frac{1}{702}\left[
           \begin{array}{cc}
             a_{11}(z)& a_{12}(z) \\
              a_{21}(z) &  a_{22}(z) \\
           \end{array}
         \right],
\]where
\[
\begin{small}
\begin{aligned}
a_{11}(z)&= ( 11-14\sqrt {17} ) {z}^{2}+ ( 29+8\sqrt {17}
 ) z+234+(85-16\sqrt {17})z^{-1}-(17+2\sqrt {17})z^{-2}
;\\
a_{12}(z)&=( 5\sqrt {17}-16 ) {z}^{3}+ (2+\sqrt {17} )
{z}^{2}+238-11\sqrt {17}+(136+29\sqrt {17})z^{-1}
;\\
a_{21}(z)&=  ( 136+29\sqrt {17} ) {z}^{2}+ ( 238-11\sqrt {17}
 ) z+(2+\sqrt {17})z^{-1}+(5\sqrt {17}-16)z^{-2};
\\
a_{22}(z)&=( -17-2\sqrt {17} ) {z}^{3}+ ( 85-16\sqrt {17} )
{z}^{2}+234z+29+8\sqrt {17}+(11-14\sqrt {17})z^{-1}.
\end{aligned}
\end{small}
\]This low-pass filter $\pa_0$ does not satisfy \eqref{mask:sym}.
However, we can employ  a very simple  orthogonal transform $E:=
{\tiny \frac{1}{\sqrt{2}}\left[
      \begin{array}{cc}
        1 & 1 \\
        1 & -1 \\
      \end{array}
\right]}$ to $\pa_0$ so that the symmetry in \eqref{mask:sym} holds.
That is, for $\wt\pa_0(z):=E\pa_0(z) E$, it is easy to verify that
$\wt\pa_0$  satisfies \eqref{mask:sym} with $c_1=c_2=1/2$ and
$\varepsilon_1=1,\varepsilon_2=-1$. Construct $\pP_{\wt
\pa_0}:=[\wt\pa_{0;0},\wt\pa_{0;1},\wt\pa_{0;2}]$ from $\wt \pa_0$.
The  matrix $\pU$ constructed by Lemma \ref{lemma:symPa} from
$\pP_{\wt \pa_0}$ is  given by:
\[
\pU=\frac{1}{\sqrt{2}}\left[
     \begin{array}{cccccc}
       z^{-1} & 0 & z^{-1} & 0 & 0 & 0 \\
       0 & z^{-1} & 0 & z^{-1} & 0 & 0 \\
       1 & 0 & -1 & 0 & 0 & 0 \\
       0 & 1 & 0 & -1 & 0 & 0 \\
       0 & 0 & 0 & 0 & \sqrt{2} & 0 \\
       0 & 0 & 0 & 0 & 0 & \sqrt{2} \\
     \end{array}
   \right].
\]
Then $\pP:=\pP_{\wt a_0}\pU$ satisfies
$\sym\pP=[z^{-1},-z^{-1}]^T[1,-1,-1,1,1,-1]$  and is given by
\[
\begin{small}
\pP=c\left[
 \begin{array}{cccccc}
234(1+\frac1z)&t_{12}(1-\frac1z)&t_{13}(1-\frac1z)&0&117\sqrt{2}(1+\frac1z)&t_{16}(1-\frac1z)
\\ \noalign{\vspace{0.05in}}
t_{21}(1-\frac1z)&t_{22}(1+\frac1z)&t_{23}(1+\frac1z)&t_{24}(1-\frac1z)&t_{25}(1-\frac1z)&t_{26}(1+\frac1z)
\end{array}
\right],
\end{small}
\]
where $c=\frac{\sqrt{6}}{1404}$ and $t_{jk}$'s are constants defined
as follows:
\[
\begin{small}
\begin{aligned}
t_{12}&=3(11-\sqrt{17});&
t_{13}&=3(\sqrt{17}-89);&
t_{16}&=15\sqrt {2}(2+\sqrt {17} );&\\
t_{21}&=13(\sqrt{17}-17);&
t_{22}&=6(2+\sqrt{17});&
t_{23}&=6(37-\sqrt{17});&\\
t_{24}&=-13(1+\sqrt{17});&
t_{25}&=-13\sqrt{2}(8+\sqrt {17});&
t_{26}&=-3\sqrt{2}(7+10\sqrt{17}).
\end{aligned}
\end{small}
\]
Applying Algorithm \ref{alg:1} to $\pP$,  we  obtain a desired
paraunitary matrix $\pP_e$ as follows:
\[
\begin{small}
\pP_e=c\left[
 \begin{array}{cccccc}
234(1+\tfrac1z)&t_{12}(1-\tfrac1z)&t_{13}(1-\tfrac1z)&0&117\sqrt{2}(1+\tfrac1z)&t_{16}(1-\tfrac1z)
\\ \noalign{\vspace{0.05in}}
t_{21}(1-\tfrac1z)&t_{22}(1+\tfrac1z)&t_{23}(1+\tfrac1z)&t_{24}(1-\tfrac1z)&t_{25}(1-\tfrac1z)&t_{26}(1+\tfrac1z)\\
\hline
\noalign{\vspace{-0.09in}}\\
t_{31}(1-\tfrac1z)&t_{32}(1+\tfrac1z)&t_{33}(1+\tfrac1z)&t_{34}(1-\tfrac1z)&t_{35}(1-\tfrac1z)&t_{36}(1+\tfrac1z)
\\ \noalign{\vspace{0.05in}}
t_{41}(1+\tfrac1z)&t_{42}(1-\tfrac1z)&t_{43}(1-\tfrac1z)&t_{44}(1+\tfrac1z)&-\sqrt{2}t_{41}(1+\tfrac1z)&t_{46}(1-\tfrac1z)\\
\hline
\noalign{\vspace{-0.09in}}\\
\frac{2}{\sqrt{3}}t_{44}&0&0&-2\sqrt{3}t_{41}&-\frac{4}{\sqrt{6}}t_{44}&0\\
0&t_{62}&t_{63}&0&0&t_{66}\\
\end{array}
\right],
\end{small}
\]
where all $t_{jk}$'s are constants given by:
\[
\begin{small}
\begin{aligned}
t_{31}&=-\sqrt{26}(61+25\sqrt {17})/4;&
t_{32}&=-3\sqrt{26}(397+23\sqrt {17})/52;\\
t_{33}&=3\sqrt{26}(553+23\sqrt {17})/52;&
t_{34}&=25\sqrt{26}(1+\sqrt { 17})/4;\\
t_{35}&=\sqrt{13}( 25\sqrt {17}-43 )/2;&
t_{36}&=15\sqrt{13}( 23\sqrt {17}-19)/26\\
t_{41}&=9\sqrt{26}(1-3\sqrt {17})/4;&
t_{42}&=-3\sqrt{26}(383+29\sqrt {17})/52;\\
%\end{aligned}
%\]
%\[
%\begin{aligned}
%
t_{43}&=3\sqrt{26}(29\sqrt {17}+227)/52;&
t_{44}&=27\sqrt{26}(1+\sqrt {17})/4;\\
%
%t_{45}&=9\sqrt{13}( 3\sqrt{17}-1)/2;&
%
t_{46}&=3\sqrt{13}(145\sqrt {17}-61)/26;&
%
%t_{51}&=9\sqrt {78} ( 1+\sqrt {17} )/2;\\
%
%t_{54}&=9\sqrt {78}(3\sqrt {17}-1 )/2;\\
%
%t_{55}&=-9\sqrt {39} ( 1+\sqrt {17} );&
%
t_{62}&=9\sqrt{78}(41\sqrt{17}-9)/26;\\
t_{63}&=9\sqrt{78}(11\sqrt{17}+9)/26;&
t_{66}&=27\sqrt{3}(\sqrt{17}+15)/\sqrt{13}.
\end{aligned}
\end{small}
\]
Note that $\pP_e$ satisfies
$\sym\pP_e=[z^{-1},-z^{-1},-z^{-1},z^{-1},1,-1]^T[1,-1,-1,1,1,-1]$
and  we have $\cs([\pP_e]_{:,j})\subseteq\cs([\pP]_{:,j})$ for all
$1\le j\le 6$. From the polyphase matrix $\PP:=\pP_e \pU^*$, we
derive two high-pass filters $\wt\pa_1, \wt \pa_2$ as follows:
\[
%\begin{small}
\begin{aligned}
\wt \pa_1(z)&=\frac{\sqrt{26}}{36504}\left[
           \begin{array}{cc}
             a^1_{11}(z)-za^1_{11}(z^{-1}) & a^1_{12}(z)+z a^1_{12}(z^{-1}) \\
             \noalign{\vspace{-0.1in}}\\
              a^1_{21}(z)+za^1_{21}(z^{-1}) &  a^1_{22}(z)-z a^1_{22}(z^{-1}) \\
           \end{array}
         \right],\\
\wt \pa_2(z)&=\;\;\frac{\sqrt{78}}{4056}\left[
           \begin{array}{cc}
             a^2_{11}(z) & a^2_{12}(z) \\
             \noalign{\vspace{-0.1in}}\\
              a^2_{21}(z) &  a^2_{22}(z)\\
           \end{array}
         \right],
\end{aligned}
%\end{small}
\]where
\[
\begin{small}
\begin{aligned}
a^1_{11}(z)&=( 433-128\sqrt {17} ) {z}^{3}+
13(25\sqrt{17}-43){z}^{2}-(1226+197\sqrt {17} ) z
;\\
a^1_{12}(z)&=( 128\sqrt {17}-433 ) {z}^{3}+ 15(23 \sqrt {17} -19 )
{z}^{2}- (758+197\sqrt {17})z
;\\
a^1_{21}(z)&= 3(133-44\sqrt {17} ) {z}^{3}+ 117( 3\sqrt {17}-1)
{z}^{2}- 3(73 \sqrt {17}+94 )z;
\\
a^1_{22}(z)&=3( 44\sqrt {17}-133) {z}^{3}+ 3(145 \sqrt {17}-61 )
{z}^{2}-3( 250+73\sqrt {17} ) z;\\
%
%
%mask 2
a^2_{11}(z)&=13( 1+\sqrt {17} )(z^3-2z^2+z);\\
a^2_{12}(z)&=13(3\sqrt {17}-1)(z^3-z);\\
a^2_{21}(z)&=( 9+11\sqrt {17} )( z^3-z );\\
a^2_{22}(z)&=( 41\sqrt {17}-9)( z^3+24z^2/137+18\sqrt {17}z^2/137+z
).
\end{aligned}
\end{small}
\]
Then the high-pass filters $\wt \pa_1$ and $\wt \pa_2$ satisfy
\eqref{highpass:sym} with $c^1_1=c^1_2=1/2$, $\varepsilon^1_1=-1,
\varepsilon^1_2=1$ and $c^2_1=c^2_2=3/2$, $\varepsilon^1_1=1,
\varepsilon^1_2=-1$, respectively.

Let $\pa_1,\pa_2$ be two high-pass filters constructed from $\wt
\pa_1, \wt \pa_2$ by $\pa_1(z):=E\wt\pa_1(z)E$ and
$\pa_2(z):=E\wt\pa_2(z)E$. Then due to the orthogonality of $E$,
$\{\pa_0,
 \pa_1,  \pa_2\}$ still forms a $\df$-band  filter
bank with the perfect reconstruction property but their symmetry
patterns are different to those of $\wt \pa_0,\wt \pa_1,\wt \pa_2$.
}
\end{example}

\section{Proofs of Theorems~\ref{thm:main:1} and \ref{thm:main:2}}In this section, we  shall prove Theorems \ref{thm:main:1} and   \ref{thm:main:2}.
The key ingredient is to prove that the coefficient supports of
$\pA_1, \ldots, \pA_J$ constructed in Algorithm \ref{alg:1} are all
contained inside $[-1,1]$. Note that each $\pA_j$ takes the form
$\pA_j=(\pB_1\cdots\pB_r)\pB_{(-k,k)}\pB_{\pQ_1}$. We first show
that the coefficient support of
$\pB:=(\pB_1\cdots\pB_r)\pB_{(-k,k)}$ is contained inside $[-1,1]$
and then show that the coefficient support of $\pB\pB_{\pQ_1}$ is
also contained inside $[-1,1]$.

Let us first present a detailed  construction for the unitary
matrices $U_{\vf}$ and $U_{G}$ that are used in Algorithm
\ref{alg:1}. For a $1\times n$  row vector $\vf$ in $\F$ such that
$\|\vf\|\neq0$,  we define $n_{\vf}$ to be the number of nonzero
entries in $\vf$ and $\e_j:=[0,\ldots,0,1,0,\ldots,0]$ to be the
$j$th unit coordinate row vector in $\R^n$. Let $E_{\vf}$ be a
permutation matrix such that $\vf
E_{\vf}=[f_1,\ldots,f_{n_{\vf}},0,\ldots,0]$ with $f_j\neq 0$ for
$j=1,\ldots, n_{\vf}$. We define
\begin{equation}
\label{Vf} V_{\vf}:=\left\{
                      \begin{array}{ll}
                        I_n, & \hbox{if $n_{\vf}=1$;} \\
                        \frac{\bar{f_1}}{|f_1|}\left(I_n-\frac{2}{\|v_\vf\|^2}v_\vf^*v_{\vf}\right), & \hbox{if $n_{\vf}>1$,}
                      \end{array}
                    \right.
\end{equation}where
$v_\vf:=\vf-\frac{f_1}{|f_1|}\|\vf\|\e_1$. Observing that
$\|v_\vf\|^2=2\|\vf\|(\|\vf\|-|f_1|)$, we can verify that $V_\vf
V_{\vf}^*=I_n$ and $\vf E_{\vf} V_\vf =\|\vf\|\e_1$. Let
$U_{\vf}:=E_{\vf}V_{\vf}$. Then $U_\vf$ is unitary and satisfies
$U_{\vf}=[\frac{\vf^*}{\|\vf\|},F^*]$ for some $(n-1)\times n$
matrix $F$ in $\F$ such that $\vf U_{\vf}=[\|\vf\|,0,\ldots,0]$. We
also define $U_{\vf}:=I_n$ if $\vf={\bf0}$ and $U_{\vf}:=\emptyset$
if $\vf=\emptyset$. Here, $U_{\vf}$ plays  the role of reducing the
number of nonzero entries in $\vf$. More generally, for an $r\times
n$ nonzero matrix $G$ of rank $m$ in $\F$, employing the above
procedure to each row of $G$, we can obtain an $n\times n$ unitary
matrix $U_{G}$ such that $GU_{G}=[R,{\bf0}]$ for some  $r \times m$
lower triangular matrix $R$ of rank $m$. If $G_1G_1^*=G_2G_2^*$,
then the above procedure produces two matrices $U_{G_1}, U_{G_2}$
such that $G_1U_{G_1}=[R,{\bf0}]$ and $G_2U_{G_2}=[R,{\bf0}]$ for
some  lower triangular matrix $R$ of full rank. It is important to
notice  that the constructions of $U_{\vf}$ and  $U_{G}$ only
involve the nonzero entries of $\vf$ and nonzero columns of $G$,
respectively. In other words, we have
\begin{equation}
\label{eq:U_f}
\begin{array}{ll}
 \, [U_{\vf}]_{j,:}=
([U_{\vf}]_{:,j})^T=\e_j,&\hbox{if $[\vf]_{j}=0$},
\\
\,[U_{G}]_{j,:}=([U_{G}]_{:,j})^T=\e_j,&\hbox{if
$[G]_{:,j}={\bf0}$.}
\end{array}
\end{equation}
%Suppose $\pQ$ has symmetry $\sym\pQ=[{\bf1}_{r_1},
%-{\bf1}_{r_2},{z}{\bf1}_{r_3 }, -z{\bf1}_{r_4}]^T[{\bf1}_{s_1},
%-{\bf1}_{s_2},\frac1z{\bf1}_{s_3 }, -\frac1z{\bf1}_{s_4}]$.
%
%By our construction of $U_{\vf}$ and $U_G$, appending ${\bf 0}$
%columns to $\pQ$ does not affect our construction of each
%$\pB_{\pq}$ and $\pB_{\pQ_0}$ (by simply taking the submatrix with
%respect to those non-appended ${\bf0}$-columns). Thus, appending
%${\bf0}$ columns and assigning with appropriate symmetry patterns,
%we can assume that $s_k=s_0$ for $k=1,2,3,4$.

Next, we establish the following lemma, which is needed later to
show that the coefficient support of
$(\pB_1\cdots\pB_r)\pB_{(-k,k)}$ is contained inside $[-1,1]$.
\begin{lemma}\label{lemma:structuralMatrixDegree2} Suppose $\pB$ is an ${s\times s}$ paraunitary matrix such that $\cs(\pB)\subseteq[-1,1]$ and  $\sym \pB=(\sym
\pth)^*\sym \pth$ with $\sym \pth=[{\bf1}_{s_1},
-{\bf1}_{s_2},{z^{-1}}{\bf1}_{s_3 }, -z^{-1}{\bf1}_{s_4}]$ for some
nonnegative integers $s_1,\ldots,s_4$ such that $s_1+s_2+s_3+s_4=s$.
Then the following statements hold.
\begin{itemize}
\item[{\rm(1)}] Let $\pp$ be a ${1 \times s}$ row vector of Laurent
polynomials with symmetry such that $\pp\pp^*=1$,
$\cs(\pp)=[k_1,k_2]$ with $k_2-k_1\ge 2$, and $\sym\pp=\varepsilon
z^{c}\sym\pth$ for some $\varepsilon\in\{-1,1\}$ and $c\in\{0,1\}$.
Let $\pq:=\pp\pB$. If $\cs(\pq)=\cs(\pp)$, then
$\cs(\pB\pB_{\pq})\subseteq[-1,1]$, where  $\pB_{\pq}$ is
constructed with respect to $\pq$ as in section~2.

\item[{\rm(2)}] Let $\pp_1,\pp_2$ be two $1\times s$ row vectors of
Laurent polynomials with symmetry such that
$\pp_{j_1}\pp_{j_2}^*=\delta(j_1-j_2)$ for $j_1,j_2=1,2$,
$\sym\pp_1=\gep_1\sym\pth$ and $\sym\pp_2=\gep_2z\sym\pth$ for some
$\gep_1,\gep_2\in\{-1,1\}$, and
$\cs(\pp_1)=\cs(\pp_2)\subseteq[-k,k]$ with $k\ge1$. Let
$\pq_1:=\pp_1\pB$ and $\pq_2:=\pp_2\pB$. If $\cs(\pq_1)=[-k,k-1]$
and $\cs(\pq_2)=[-k+1,k]$, then
$\cs(\pB\pB_{(\pq_1,\pq_2)})\subseteq[-1,1]$, where
$\pB_{(\pq_1,\pq_2)}$ is constructed with respect to the pair
$(\pq_1,\pq_2)$ as in section~2.
\end{itemize}
\end{lemma}
\begin{proof}
Due to $\sym\pp=\varepsilon z^{c}\sym\pth$, as we discussed  in
section~2, there is an $\pU_{\pp,\gep}$ such that
$\pp\pU_{\pp,\gep}$ takes the form in \eqref{eq:type1}.  Since
$\pU_{\pp,\gep}$ is a product of a permutation matrix and a diagonal
matrix of monomials, we shall consider the case that
$\pU_{\pp,\gep}=I_s$, while the proofs for other cases of
$\pU_{\pp,\gep}$ can be obtained accordingly. Then $\pp$ takes the
standard form in \eqref{eq:type1} with $\vf_1\neq {\bf0}$. In this
case, $s_1>0$ and $s_2>0$ due to $\|\vf_1\|=\|\vf_2\|\neq0$. By our
assumptions, $\pq:=\pp\pB$ must take the following form:
\[
\begin{small}
\begin{aligned}
\pq:=\pp\pB=&[
        \wt\vf_1,
       -\wt\vf_2,
 \wt\vg_1,
      - \wt\vg_2
   ]z^{k_1} +
  [      \wt \vf_3,
       -\wt\vf_4,
      \wt\vg_3,
       -\wt\vg_4
    ]z^{k_1+1}
   +\sum_{n=k_1+2}^{k_2-2}\coeff(\pp\pB,n)z^n
  \\&+
 [
       \wt\vf_3,
       \wt\vf_4,
       \wt\vg_1,
      \wt \vg_2
  ]z^{k_2-1} +
  [
      \wt\vf_1 ,
       \wt\vf_2,
  {\bf0},
      {\bf0}]z^{k_2}
      \end{aligned}
\end{small}
\]
with $\wt\vf_1\neq{\bf0}$. Then $\pB_{\pq}$ is given by
\eqref{eq:MatrixForvectorDegBy2} with  $\vf_1$, $\vf_2$, $\vg_1$,
$\vg_2$, $F_1$, $F_2$, $G_1$, $G_2$ being replaced by $\wt \vf_1$,
$\wt\vf_2$, $\wt\vg_1$, $\wt\vg_2$, $\wt F_1$, $\wt F_2$, $\wt G_1$,
$\wt G_2$ respectively and
%\[
%\pB^*_{\pq}:=\frac{1}{c}\left[
% \begin{array}{c|c|c|c}
% \wt\vf_1(z+\frac{{c_0}}{c_{\wt\vf_1}}+\frac1z)& \wt\vf_2(z-\frac1z) & \wt\vg_1(1+\frac1z) &\wt \vg_2(1-\frac1z) \\
%\wt F_1&  {\bf 0} &{\bf 0} &{\bf 0}\\
%\hline
%-\wt\vf_1(z-\frac1z) & -\wt\vf_2(z-\frac{{c_0}}{c_{\wt\vf_2}}+\frac1z) & -\wt\vg_1(1-\frac1z) & -\wt\vg_2(1+\frac1z) \\
%{\bf 0}& \wt F_2  &{\bf 0} &{\bf 0}\\
%\hline
%\frac{c_{\wt\vg_1}}{c_{\wt\vf_1}}\wt\vf_1(1+z) & -\frac{c_{\wt\vg_1}}{c_{\wt\vf_2}}\wt\vf_2(1-z) & c_{\wt\vg_1'}\wt\vg_1' & {\bf 0}\\
%{\bf 0}&  {\bf 0} & \wt G_1&{\bf 0}\\
%\hline
%\frac{c_{\wt\vg_2}}{c_{\wt\vf_1}}\wt\vf_1(1-z) & -\frac{c_{\wt\vg_2}}{c_{\wt\vf_2}}\wt\vf_2(1+z) & {\bf 0} & c_{\wt\vg_2'} \wt\vg_2'\\
%{\bf 0}&  {\bf 0} &{\bf 0} &\wt G_2\\
%\end{array}
%\right]
%\]
all constants $c_{\wt\vf_1},c_{\wt \vg_1}, c_{\wt \vg_2}, c_0, c,
c_{\wt \vg_1'}, c_{\wt\vg_2'}$ being defined accordingly.

Also, due to the symmetry pattern and $\cs(\pB)\subseteq[-1,1]$,
$\pB$ is of the form:
\begin{equation}
\label{eq:StructralMatrixFormGeneral}
\begin{small}
\pB=\left[
  \begin{array}{c|c|c|c}
    A_1(z+\frac1z)+D_1 & A_3(z-\frac1z) & B_3(1+\frac1z) & B_4(1-\frac1z) \\
    \hline
\noalign{\vspace{-0.1in}}&&&\\
    A_2(z-\frac1z) & A_4(z+\frac1z)+D_2 & C_3(1-\frac1z) & C_4(1+\frac1z) \\
\hline
\noalign{\vspace{-0.1in}}&&&\\
    B_1(1+z) & C_1(1-z)& A_5(z+\frac1z)+D_3 & A_7(z-\frac1z) \\
\hline
\noalign{\vspace{-0.1in}}&&&\\
    B_2(1-z) & C_2(1+z) & A_6(z-\frac1z) & A_8(z+\frac1z)+D_4 \\
  \end{array}
\right],
\end{small}
\end{equation}
where $A_j$'s, $B_j$'s, $C_j$'s and $D_j$'s are all constant
matrices in $\F$ and $D_j$ is  of size $s_j\times s_j$ for
$j=1,\ldots,4$.

Let
$\mathcal{I}:=\{1,s_1+1,(1-\delta(s_3))(s_1+s_2+1),(1-\delta(s_4))(s_1+s_2+s_3+1)\}$
be an index set. It is easy to verify that
$\cs([\pB\pB_{\pq}]_{:,j})\subseteq[-1,1]$ for all
$j\notin\mathcal{I}$. Hence, by $\cs(\pB\pB_{\pq})\subseteq[-2,2]$,
we only need to compute $\coeff([\pB\pB_{\pq}]_{:,j},2)$ and
$\coeff([\pB\pB_{\pq}]_{:,j},-2)$ for those $j\in\mathcal{I}$. Let
us show that $\coeff([\pB\pB_{\pq}]_{:,j},2)={\bf0}$ for $j=1$,
i.e., the coefficient vector of $z^2$ for the first column of
$\pB\pB_{\pq}$ is $\bf0$.  By $\coeff(\pp\pB, {k_1}) =
\coeff(\pp,{k_1+1})\coeff(\pB,{-1})+\coeff(\pp,{k_1})\coeff(\pB,0)$,
we have
\begin{equation}
\label{eq:f1f2g1g2}
\begin{small}
\begin{aligned}
\wt\vf_1 &=\vf_3A_1+\vf_4A_2+\vf_1D_1+\vg_1B_1-\vg_2B_2;\\
\wt\vf_2&=\vf_3A_3+\vf_4A_4+\vf_2D_2-\vg_1C_1+\vg_2C_2;\\
\wt\vg_1&= \vf_3B_3+\vf_4C_3+\vg_3A_5+\vg_4A_6+\vf_1B_3-\vf_2C_3+\vg_1D_3;\\
\wt\vg_2&=\vf_3B_4+\vf_4C_4+\vg_3A_7+\vg_4A_8-\vf_1B_4+\vf_2C_4+\vg_2D_4.
\end{aligned}
\end{small}
\end{equation}
Similarly, by
$\coeff(\pB\pB_{\pq},2)=\coeff(\pB,1)\coeff(\pB_{\pq},1)$, we have
\[
\begin{scriptsize}
\begin{aligned}
\coeff([\pB\pB_{\pq}]_{:,1},2) = \frac1c\left[
  \begin{array}{cccc}
    A_1&A_3&{\bf0}&{\bf0} \\
    A_2 & A_4& {\bf0}&{\bf0}\\
    B_1 & -C_1 & A_5 & A_7 \\
    -B_2 & C_2 & A_6 & A_8 \\
  \end{array}
\right] \left[
  \begin{array}{c}
    \wt\vf_1^* \\
    -\wt\vf_2^* \\
    \wt\vg_1^* \\
    -\wt\vg_2^* \\
  \end{array}
\right]=\frac1c\left[
  \begin{array}{c}
    A_1\wt\vf_1^*-A_3\wt\vf_2^* \\
    A_2\wt\vf_1^*-A_4\wt\vf_2^* \\
   B_1\wt\vf_1^*+C_1\wt\vf_2^*+A_5\wt\vg_1^*-A_7\wt\vg_2^*  \\
    -B_2\wt\vf_1^*-C_1\wt\vf_2^*+A_6\wt\vg_1^*-A_8\wt\vg_2^* \\
  \end{array}
\right].
\end{aligned}
\end{scriptsize}
\]
Due to $\pB\pB^*=I_{s}$, we obtain
\[
\begin{small}
 \left\{
  \begin{array}{l}
    A_1A_1^*-A_3A_3^*={\bf0},\, A_1A_2^*-A_3A_4^*={\bf0};\\
    A_1D_1^*+D_1A_1^*+B_3B_3^*-B_4B_4^*={\bf0}; \\
    D_1A_2^*-A_3D_2^*+B_3C_3^*-B_4C_4^*={\bf0};\\
    A_1B_1^*+A_3C_1^*+B_3A_5^*-B_4A_7^*={\bf0}; \\
    -A_1B_2^*-A_3C_2^*+B_3A_6^*-B_4A_8^*={\bf0}.\\
  \end{array}
\right.
\end{small}
\]
Applying the above identities to $ A_1\wt\vf_1^*-A_3\wt\vf_2^*$ and
using \eqref{eq:f1f2g1g2}, we get
\begin{small}
\[
\begin{aligned}
A_1\wt\vf_1^*-A_3\wt\vf_2^* &=
A_1(\vf_3A_1+\vf_4A_2+\vf_1D_1+\vg_1B_1-\vg_2B_2)^*
  \\&\quad-A_3(\vf_3A_3+\vf_4A_4+\vf_2D_2-\vg_1C_1+\vg_2C_2)^*
  \\&=(A_1A_1^*-A_3A_3^*)\vf_3^*
  +(A_1A_2^*-A_3A_4^*)\vf_4^*
  +(A_1D_1^*)\vf_1^*
  \\&\quad+(-A_3D_2^*)\vf_2^*
 +(A_1B_1^*+A_3C_1^*)\vg_1^*
  -(A_1B_2^*+A_3C_2^*)\vg_2^*\\
\end{aligned}
\]
\[
\begin{aligned}
\qquad\qquad\qquad\quad&=-(D_1A_1^*+B_3B_3^*-B_4B_4^*)\vf_1^*
  -(D_1A_2^*+B_3C_3^*-B_4C_4^*)\vf_2^*
 \\&\quad-(B_3A_5^*-B_4A_7^*)\vg_1^*
  -(B_3A_6^*-B_4A_8^*)\vg_2^*
  \\&= -D_1(\vf_1A_1+\vf_2A_2)^*-B_3(\vf_1B_3+\vf_2C_3+\vg_1A_5+\vg_2A_6)^*
   \\&\quad+B_4(\vf_1B_4+\vf_2C_4+\vg_1A_7+\vg_2A_8)^*={\bf0},
\end{aligned}
\]
\end{small}where the last above identity follows by
$\coeff(\pp\pB,{k_2+1})=\coeff(\pp\pB,{k_1-1})={\bf0}$. Similarly,
we can show that  $A_2\wt\vf_1^*-A_4\wt\vf_2^*={\bf0}$,
$B_1\wt\vf_1^*+C_1\wt\vf_2^*+A_5\wt\vg_1^*-A_7\wt\vg_2^*={\bf0}$,
and
$-B_2\wt\vf_1^*-C_1\wt\vf_2^*+A_6\wt\vg_1^*-A_8\wt\vg_2^*={\bf0}$.
Hence, $\coeff([\pB\pB_{\pq}]_{:,1},2)={\bf0}$. By similar
computations as above and using the paraunitary property of $\pB$,
we have $\coeff([\pB\pB_{\pq}]_{:,j},\pm2)={\bf0}$ for all
$j\in\mathcal{I}$. Therefore, we conclude that
$\cs(\pB\pB_{\pq})\subseteq[-1,1]$.  Item (1) holds.

For item (2), up to a permutation matrix $E_{(\pq_1,\pq_2)}$ as in
section~2,  $\pB_{(\pq_1,\pq_2)}$ takes the form in
\eqref{eq:MatrixForvectorDegBy2SpecialCase}. Since $\pB$ takes the
form in \eqref{eq:StructralMatrixFormGeneral}, to show that the
coefficient support of $\pB\pB_{(-k,k)}$ is contained inside
$[-1,1]$, we need to show that all the coefficient vectors
$A_1\wt\vg_1^*-A_3\wt\vg_2^*$, $A_2\wt\vg_1^*-A_4\wt\vg_2^*$,
$A_5\wt\vg_3^*-A_7\wt\vg_4^*$, and $A_6\wt\vg_3^*-A_8\wt\vg_4^*$ are
 zero. Again, using the paraunitary property of $\pB$ and
expressing $\wt\vg_1,\wt\vg_2,\wt\vg_3,\wt\vg_4$ in terms of the
original vectors from $\pp_1,\pp_2$ similar to \eqref{eq:f1f2g1g2},
we conclude that $\cs(\pB\pB_{(\pq_1,\pq_2)})\subseteq[-1,1]$.
\end{proof}

With the result of Lemma \ref{lemma:structuralMatrixDegree2}, the
next lemma shows that the coefficient support of
$\pB:=(\pB_1\cdots\pB_r)\pB_{(-k,k)}$ is contained inside $[-1,1]$.
Moreover, the next lemma  shows that the coefficient support of
$\pA:=\pB\pB_{\pQ_1}$ is also contained inside $[-1,1]$.

\begin{lemma}
\label{lemma:MultiColumnMatrixReducedBY2} Suppose $\pQ$ is an
$r\times s$ matrix of Laurent polynomials  such that $\pQ\pQ^*=I_r$,
$\sym \pQ$ satisfies \eqref{Q:sym:standard},
%=
%[{\bf1}_{r_1},-{\bf1}_{r_2},z{\bf1}_{r_3},-z{\bf1}_{r_4}]^T[
%{\bf1}_{s_1}, -{\bf1}_{s_2}, z^{-1}{\bf1}_{s_3 },
%-z^{-1}{\bf1}_{s_4}]$
and
 $\cs(\pQ)=[k_1,k_2]$ with $k_2-k_1\ge 1$. Then there exists an
$s\times s$ paraunitary matrix $\pA$ of Laurent polynomials with
symmetry such that
\begin{itemize}
\item[{\rm(1)}] $\cs(\pA) \subseteq [-1, 1]$ and
$|\cs(\pQ\pA)|\le|\cs(\pQ)|-|\cs(\pA)|$;

\item[{\rm(2)}] if the $j$th column $\pp:=[\pQ]_{:,j}$ of $\pQ$ satisfies $\coeff(\pp,k_1)=\coeff(\pp,k_2)={\bf0}$, then
 $[\pA]_{j,:}=([\pA]_{:,j})^T=\e_j$. That is, any entry in the $j$th
row or $j$th column of $\pA$ is zero except that  the $(j,j)$-entry
$[\pA]_{j,j}=1$;

\item[{\rm(3)}] $\sym \pA=[
{\bf1}_{s_1}, -{\bf1}_{s_2}, z{\bf1}_{s_3 }, -z{\bf1}_{s_4}]^T[
{\bf1}_{s_1'}, -{\bf1}_{s_2'}, z^{-1}{\bf1}_{s_3'},
-z^{-1}{\bf1}_{s_4'}]$ for some nonnegative integers
$s_1',\ldots,s_4'$ such that $s_1'+s_2'+s_3'+s_4'=s$.
\end{itemize}
\end{lemma}

\begin{proof}
Let $\pA=(\pB_1\cdots\pB_r)\pB_{(-k,k)}\pB_{\pQ_1}$ be constructed
as in Algorithm \ref{alg:1}, where
$\pQ_1:=\pQ(\pB_1\cdots\pB_r)\pB_{(-k,k)}$, $\pB_{(-k,k)}$ is
constructed in the inner \texttt{while} loop of Algorithm
\ref{alg:1}, and $\pB_1,\ldots,\pB_r$ is constructed in the
\texttt{for} loop of Algorithm \ref{alg:1}. If $k_2\neq-k_1$, then
$\pB_1=\cdots=\pB_r=\pB_{(-k,k)}=I_s$ and $\pA$ is simply
$\pB_{\pQ_1}$, where $\pQ_1=\pQ$  is of the form in
\eqref{eq:Q0:form} with either $\coeff(\pQ_1,-k)={\bf0}$ or
$\coeff(\pQ_1,k)={\bf0}$. In this case, by the  construction of
$\pB_{\pQ_1}$ as in section~2, all items in Lemma
\ref{lemma:MultiColumnMatrixReducedBY2} hold. We are already done.
So, without loss of generality, we assume that $k_2=-k_1=k$.

We first show that the coefficient support of $\pB_1\cdots\pB_r$ is
contained inside $[-1,1]$. Let $\pp_j:=[\pQ]_{j,:}$, $\pB_0:=I_s$,
and $\pq_j:=\pp_j\pB_{0}\cdots\pB_{j-1}$ for $j=1,\ldots,r$. Suppose
we already show that $\cs(\pB_{0}\cdots\pB_{j-1})\subseteq[-1,1]$
for $j\ge 1$. Then, according to Algorithm \ref{alg:1},
$\pB_j=\pB_{\pq_j}$ if $\cs(\pp_j)=\cs(\pq_j)$, $|\cs(\pq_j)|\ge 2$,
and one of $\coeff(\pq_j,k)$ and $\coeff(\pq_j,-k)$ is nonzero;
otherwise $\pB_j=I_{s}$. Note that $\pB_{0}\cdots\pB_{j-1}$ is
paraunitary and satisfies
$\sym(\pB_{0}\cdots\pB_{j-1})=(\sym\pth)^*\sym\pth$ with $\sym
\pth=[{\bf1}_{s_1}, -{\bf1}_{s_2},{z^{-1}}{\bf1}_{s_3 },
-z^{-1}{\bf1}_{s_4}]$. By item (1) of Lemma
\ref{lemma:structuralMatrixDegree2}, the coefficient support of
$\pB_{0}\cdots\pB_{j-1}\pB_{j}$ is also contained inside  $[-1,1]$.
By induction, the coefficient support of $\pB_1\cdots\pB_r$ is
contained inside $[-1,1]$. Moreover, $\pB_1\cdots\pB_r$ takes the
form in \eqref{eq:StructralMatrixFormGeneral}. Next, since
$\pB_{(-k,k)}$ is constructed recursively from pairs $(\pq_1,\pq_2)$
of $\pQ_0:=\pQ(\pB_1\cdots\pB_r)$, by applying induction again and
using item (2) of Lemma \ref{lemma:structuralMatrixDegree2}, we
conclude that the coefficient support  of
$\pB:=(\pB_1\cdots\pB_r)\pB_{(-k,k)}$ is contained inside $[-1,1]$.

Due to the Property (P1), (P2) of $\pB_{\pq}$ and (P3), (P4) of
$\pB_{(\pq_1,\pq_2)}$, $\pB_1,\ldots,\pB_r$ and $\pB_{(-k,k)}$
reduce $\pQ$  of the form in \eqref{eq:P:Standardform} to
$\pQ_1=\pQ(\pB_1\cdots\pB_r)\pB_{(-k,k)}=\pQ\pB$ of the form in
\eqref{eq:Q0:form} with at least one of $\coeff(\pQ_1,-k)$ and
$\coeff(\pQ_1,k)$ being ${\bf0}$. As constructed in section~2,
$\pB_{\pQ_1}=I_s$ for the case that
$\coeff(\pQ_1,-k)=\coeff(\pQ_1,k)={\bf0}$, or
$\pB_{\pQ_1}=\diag(U_1\pW_1,I_{s_3+s_4})E$ for the case
$\coeff(\pQ_1,k)\neq{\bf0}$, or $\pB_{\pQ_1}:=\diag(I_{s_1+s_2},
U_3\pW_3)E$ for the case that $\coeff(\pQ_1,-k)\neq{\bf0}$. We next
show that $\cs(\pB\pB_{\pQ_1})\subseteq[-1,1]$.

Let $\pQ$ take the form in \eqref{eq:P:Standardform} and $\pQ_1$
take the form in \eqref{eq:Q0:form} with
$\coeff(\pQ_1,k)\neq{\bf0}$. Then
$\pB_{\pQ_1}:=\diag(U_1\pW_1,I_{s_3+s_4})E$ with $U_1$, $\pW_1$, and
$E$ being constructed  as in section~2. Note that $\pB$ takes the
form in \eqref{eq:StructralMatrixFormGeneral}.
%We only need to show
%that $\cs(\pB \diag(U_1\pW_1,I_{s_3+s_4}))\subseteq[-1,1]$.
Define
\[
\begin{small}
%\begin{aligned}
\, [G_1,G_2,F_3,F_4,G_5,G_6,F_7,F_8]:=\left[
  \begin{array}{cccccccc}
     G_{11} & G_{21}&F_{31} & F_{41} & G_{51} & G_{61}&F_{71} & F_{81} \\
    G_{12} & G_{22} &F_{32} & F_{42}  &  G_{52} & G_{62} &F_{72} & F_{82}  \\
  \end{array}
\right]
%,
%\\ [G_5,G_6,F_7,F_8]&:=\left[
%   \begin{array}{cccc}
%        G_{51} & G_{61}&F_{71} & F_{81} \\
%    G_{52} & G_{62} &F_{72} & F_{82}   \\
%     \end{array}
%   \right]
.
%\end{aligned}
\end{small}
\]
By $\coeff(\pQ_1,k)= \coeff(\pQ,{k-1})\coeff(\pB,1
)+\coeff(\pQ,{k})\coeff(\pB,0)$, we have
\begin{equation}\label{eq:wtG1wtG2}
\begin{small}
\begin{aligned}
        \wt G_1&=G_5A_1+G_6A_2+F_7B_1-F_8B_2+G_1 D_1+F_3B_1+F_4 B_2;\\
        \wt G_2&=G_5A_3+G_6A_4-F_7C_1+F_8C_2+G_2D_2+F_3C_1+F_4C_2;\\
        {\bf0}&=F_7A_5+F_8A_6+G_1B_3+G_2C_3+F_3D_3=:\wt F_3;\\
        {\bf0}&=F_7A_7+F_8A_8+G_1B_4+G_2C_4+F_4D_4=:\wt F_4,\\
\end{aligned}
\end{small}
\end{equation}where $\wt G_1,\wt G_2$ are matrices defined in \eqref{eq:wtG}.
Then $U_1=\diag(U_{\wt G_1}, U_{\wt G_2})$ and $\pW_1$ is defined as
in \eqref{eq:W1W3}.
%\[
%\pW_1=\left[
%           \begin{array}{c|c|c|c}
%             \pU_1 & & -\pU_2&  \\
%             \hline
%             & I_{s_1-{m_1}} & &    \\
%              \hline
%             -\pU_2 & & \pU_1 &\\
%              \hline
%            & & & I_{s_2-m_1}\\
%           \end{array}
%         \right],
%\]
%where
%$\pU_1(z):=\diag(\frac{1+z^{-1}}{2},\cdots,\frac{1+z^{-1}}{2})$ is
%of size $m_1\times m_1$ with $m_1$ the rank of $\wt G_1$ and
%$\pU_2(z):=\pU_1(-z)$.
By the coefficient supports of $\pB$ and $\pB_{\pQ_1}$, we only need
to check that $\coeff(\pB\diag(U_1\pW_1,I_{s_3+s_4}),{-2})={\bf0}$.
Let $V_{11},V_{12}, V_{21}, V_{22}$ be diagonal matrices of size
$s_1\times s_1$, $s_1\times s_2$, $s_2\times s_1$, $s_2\times s_2$,
respectively, and satisfy $\diag(V_{j\ell})=[{\bf1}_{m_1},{\bf0}]$
for $j,\ell=1,2$, where $m_1$ is the rank of $\wt G_1$. Then
\[
\begin{aligned}
&\coeff(\pB\diag(U_1\pW_1,I_{s_3+s_4}),{-2})=\coeff(\pB,{-1})\cdot\coeff(\diag(U_1\pW_1,I_{s_3+s_4}),{-1})\\&=
\begin{small}
\left[
  \begin{array}{cccc}
   A_1 & -A_3 & B_3 & -B_4\\
   -A_2 & A_4 &-C_3  & C_4  \\
   {\bf0} & {\bf0} & A_5 &-A_7\\
   {\bf0} & {\bf0}& -A_6 & A_8\\
  \end{array}
\right]\left[
 \begin{array}{cccc}
   U_{\wt G_1}V_{11}& U_{\wt G_1}V_{12} & {\bf0} & {\bf0}\\
   U_{\wt G_2}V_{21} & U_{\wt G_2}V_{22} & {\bf0} & {\bf0}\\
   {\bf0} & {\bf0}& {\bf0} &{\bf0}\\
   {\bf0} & {\bf0}& {\bf0}& {\bf0}\\
  \end{array}
\right].
\end{small}
\end{aligned}
\]Thus, we need to show  ${\scriptsize A_1U_{\wt G_1}V_{1j}-A_3U_{\wt G_2}V_{2j}={\bf0}}$ and
${\scriptsize A_2U_{\wt G_1}V_{1j}-A_4U_{\wt G_2}V_{2j}={\bf0}}$,
for $j=1,2$, which is equivalent to  showing that  ${\scriptsize
V_{j1}U_{\wt G_1}^*A_1^*-V_{j2}U_{\wt G_2}^*A_3^*={\bf0}}$ and
$V_{j1}U_{\wt G_1}^*A_2^*-V_{j2}U_{\wt G_2}^*A_4^*={\bf0}$ for
$j=1,2$. Since $\wt G_1U_{\wt G_1}=[R,{\bf0}]$ and $\wt G_2 U_{\wt
G_2}=[R,{\bf0}]$, for some lower triangular matrix $R$ of full rank
$m_1$, %we have $\wt G_1=[R,{\bf0}]V_1 U_{\wt G_1}^*$ and $\wt
%G_2=[R,{\bf0}]V_2 U_{\wt G_2}^*$. Hence,
it is equivalent to proving
that ${\scriptsize \wt G_1A_1^*-\wt G_2A_3^*={\bf0}}$ and
${\scriptsize \wt G_1A_2^*-\wt G_2A_4^*={\bf0}}$. By
\eqref{eq:wtG1wtG2}, we have,
\[
\begin{small}
\begin{aligned}
\wt G_1A_1^*-\wt G_2A_3^*&= \wt G_1A_1^*-\wt G_2A_3^*+\wt
F_3B_3^*-\wt F_4B_4^*
\\&=  (G_5A_1+G_6A_2+F_7B_1-F_8B_2+G_1 D_1+F_3B_1+F_4 B_2)A_1^*\\&
  \quad-(G_5A_3+G_6A_4-F_7C_1+F_8C_2+G_2D_2+F_3C_1+F_4C_2)A_3^*\\&
  \quad+(F_7A_5+F_8A_6+G_1B_3+G_2C_3+F_3D_3)B_3^*
\\&  \quad-(F_7A_7+F_8A_8+G_1B_4+G_2C_4+F_4D_4)B_4^*\\
%\end{aligned}
%\]
%\[
%\begin{aligned}
&= G_5(A_1A_1^*-A_3A_3^*)+G_6(A_2A_1^*-A_4A_3^*) %=0
\\&\quad
+F_7(B_1A_1^*+C_1A_3^*+A_5B_3^*-A_7B_4^*) %=0
\\&\quad
+F_8(-B_2A_1^*-C_2A_3^*+A_6B_3^*-A_8B_4^*) %=0;
\\
&\quad+G_1(D_1A_1^*+B_3B_3^*-B_4B_4^*)
+G_2(-D_2A_3^*+C_3B_3^*-C_4B_4^*)
\\
&\quad+F_3(B_1A_1^*-C_1A_3^*+D_3B_3^*)
+F_4(B_2A_1^*-C_2A_3^*-D_4B_4^*)={\bf0},
\end{aligned}
\end{small}
\]
where the last identity follows from $\pB\pB^*=I_{s}$ and
$\coeff(\pQ\pB, {k+1})={\bf0}$.  Similarly, $\wt G_1A_2^*-\wt
G_2A_4^*={\bf0}$. The computation for showing
$\cs(\pB\pB_{\pQ_1})\subseteq[-1,1]$ with
$\pB_{\pQ_1}=\diag(I_{s_1+s_2},U_3\pW_3)E$ is similar. Consequently,
 $\cs(\pB\pB_{\pQ_1})\subseteq[-1,1]$. Therefore, item (1)
holds. Item (2) is due to the property \eqref{eq:U_f} of $U_{\vf}$
and $U_{G}$.

Note that $\sym\pB=(\sym\pth)^*\sym\pth$ with $\sym\pth = [
{\bf1}_{s_1}, -{\bf1}_{s_2},z^{-1}{\bf1}_{s_3 },
-z^{-1}{\bf1}_{s_4}]$. And by the construction of $\pB_{\pQ_1}$,
$\sym\pB_{\pQ_1}=(\sym\pth)^*[ {\bf1}_{s_1'}, -{\bf1}_{s_2'},
z^{-1}{\bf1}_{s_3'}, -z^{-1}{\bf1}_{s_4'}]$ for some nonnegative
integers $s_1',\ldots,s_4'$ depending on the rank of $\wt G_1$ or
$\wt G_3$ (see section~2). Consequently, item (3) holds. This also
completes the proof of Algorithm \ref{alg:1}.
\end{proof}

Now, we are ready to prove Theorems \ref{thm:main:1} and
\ref{thm:main:2}.

{\it Proof of Theorems \ref{thm:main:1} and  \ref{thm:main:2}:} The
sufficiency part of Theorem \ref{thm:main:2} is obvious. We only
need to show the necessary part. Suppose
$\sym\pP=(\sym\pth_1)^*\sym\pth_2$. Let
$\pQ:=\pU^*_{\sym\pth_1}\pP\pU_{\sym\pth_2}$ and
$\cs(\pQ):=[k_1,k_2]$. Then $\sym\pQ$ satisfies
\eqref{Q:sym:standard}. By Lemma
\ref{lemma:MultiColumnMatrixReducedBY2}, the step of support
reduction in Algorithm \ref{alg:1} produces a sequence of
paraunitary matrices $\pA_1,\ldots,\pA_J$ with coefficient support
contained inside $[-1,1]$ such that
$\pQ\pA_1\cdots\pA_J=[I_r,{\bf0}]$. Due to item (1) of Lemma
\ref{lemma:MultiColumnMatrixReducedBY2}, $J\le \lceil
\frac{k_2-k_1}{2}\rceil$. Let $\pP_j:=\pA_j^*$,
$\pP_0:=\pU_{\sym\pth_2}^*$ and
$\pP_{J+1}:=\diag(\pU_{\sym\pth_1},I_{s-r})$. Then
$\pP_e:=\pP_{J+1}\pP_J\cdots\pP_1\pP_0$ satisfies
$[I_r,{\bf0}]\pP_e=\pP$. By item (3) of
Lemma~\ref{lemma:MultiColumnMatrixReducedBY2}, $(\pP_{j+1}, \pP_j)$
has mutually compatible symmetry for all $0\le j\le J$. The claim
that $|\cs([\pP_e]_{k,j})|\le \max_{1\le n\le r}|\cs([\pP]_{n,j})|$
for $1\le j,k\le s$ follows from item (2) of Lemma
\ref{lemma:MultiColumnMatrixReducedBY2}. Hence, all claims in
Theorems \ref{thm:main:1} and \ref{thm:main:2} have been verified.
 $\endproof$

\end{document}